\documentclass[12pt]{article}

\pdfoutput=1

\usepackage[top=80pt,bottom=85pt,left=85pt,right=85pt]{geometry}
\usepackage{amsmath,amssymb,graphicx,float,color,tikz,cite,savesym,wasysym}
\usepackage{caption}
\usepackage{subcaption}
\usepackage[debug,pageanchor=false]{hyperref}
\definecolor{link}{rgb}{.8,.15,.1}
\hypersetup{colorlinks=true,linkcolor=link,citecolor=link,urlcolor=link,linktocpage}

\newcommand{\ii}{\mathrm{i}}
\newcommand{\dd}{\mathrm{d}}
\newcommand{\ee}{\mathrm{e}}

\newtheorem{theorem}{Theorem}

\makeatletter
\@addtoreset{equation}{section}
\makeatother

\usepackage[normalem]{ulem}
 
\begin{document}

\begin{titlepage}

	\begin{center}

	\vskip .5in 
	\noindent

	{\Large \bf{Leaps and bounds towards scale separation}}

	\bigskip\medskip

	 G. Bruno De Luca,$^1$  Alessandro Tomasiello$^{2}$\\

	\bigskip\medskip
	{\small 
$^1$ Stanford Institute for Theoretical Physics, Stanford University,\\
382 Via Pueblo Mall, Stanford, CA 94305, United States
\\	
	\vspace{.3cm}
$^2$ Dipartimento di Matematica, Universit\`a di Milano--Bicocca, \\ Via Cozzi 55, 20126 Milano, Italy \\ and \\ INFN, sezione di Milano--Bicocca
	
		}

	\vskip .5cm 
	{\small \tt alessandro.tomasiello@unimib.it, gbdeluca@stanford.edu}
	\vskip .9cm 
	     	{\bf Abstract }
	\vskip .1in
	\end{center}

	\noindent

In a broad class of gravity theories, the equations of motion for vacuum compactifications give a curvature bound on the Ricci tensor minus a multiple of the Hessian of the warping function. Using results in so-called Bakry--\'Emery geometry, we put rigorous general bounds on the KK scale in gravity compactifications in terms of the reduced Planck mass or the internal diameter. 
We reexamine in this light the local behavior in type IIA for the class of supersymmetric solutions most promising for scale separation. We find that the local O6-plane behavior cannot be smoothed out as in other local examples; it generically turns into a formal partially smeared O4.

	\noindent

	\vfill
	\eject

	\end{titlepage}

\tableofcontents

\section{Introduction} 
\label{sec:intro}
In the study of string compactifications, there seems to be a tension between what one can achieve by solving directly the equations of motion in ten (or eleven) dimensions, and what is suggested by effective field theory methods. This is famously the case for de Sitter vacua. But a similar issue presents itself even for negative cosmological constant: in explicit ten-dimensional solutions, $\sqrt{|\Lambda|}$ is usually of the same order as the masses $m_\mathrm{KK}$ of the KK modes. 
 
In contrast, effective field theory methods seem to lead quite naturally to  ``scale separation'' $\sqrt{|\Lambda|}\ll m_\mathrm{KK}$, and indeed rely on it. One might expect for example that many vacua that seem to be Minkowski in the supergravity approximation actually have a $\Lambda$ negative and small once stringy corrections or instantons are more carefully taken into account. Such effects are so far notoriously difficult to compute reliably, but the AdS vacua proposed in \cite{kklt} are of this type.\footnote{Even Minkowski vacua with ${\mathcal N}=1$ are a priori not protected against instanton corrections, and it was recently conjectured  \cite{palti-vafa-weigand} that such protection is either due to ${\mathcal N}>1$ or to supersymmetry in higher dimensions.} A similar strategy is to actively modify a Minkowski vacuum by introducing (additional) fluxes; in a $d=4$ effective description, these give rise to terms in the potential whose coefficients depend on the flux quanta. Starting with a Calabi--Yau compactification and taking the flux integer $N$ for $F_4$ to be large, one obtains $|\Lambda|\sim N^{-3/2}$, $m_\mathrm{KK}\sim N^{-1/4}$, and ${\mathcal N}=1$ supersymmetry \cite{dewolfe-giryavets-kachru-taylor} (see also \cite{camara-font-ibanez}). This however also involves several O6-planes, whose back-reaction is difficult to describe appropriately.\footnote{A strategy using F-theory seven-branes was proposed in \cite{polchinski-silverstein}.} We will come back to this particular class of solutions soon.

A complementary approach comes from the so-called ``swampland program'', which abstracts general lessons from the many solutions which are solidly believed to exist, in conjunction with general expectations from quantum gravity. The swampland distance conjecture \cite{ooguri-vafa-swamp} predicts a tower of states to appear at the boundary of moduli space, of masses exponential in the distance from it. Inspired by this, the $\Lambda\to 0$ limit was conjectured \cite{lust-palti-vafa} to always yield masses $m \sim |\Lambda|^\alpha$, with $\alpha>0$ of order one. A stronger version of this conjecture, also considered in \cite{lust-palti-vafa}, would be that $\alpha=1/2$ for supersymmetric solutions; this would indicate that the above AdS$_4\times \mathrm{CY}_6$ solutions of \cite{dewolfe-giryavets-kachru-taylor} are somehow invalid. 

In this paper we analyze the issue of scale separation from both points of view.
We will first derive some general bounds on the KK scale which apply to any compactification,
and then revisit the particular class in \cite{dewolfe-giryavets-kachru-taylor}; the two sets of results will be related by the role of O-planes. 

In Sec.~\ref{sec:be} we show a general bound for compactifications of any $D$-dimensional gravity theory down to $d$ dimensions, whose stress-energy tensor satisfies 
	a certain condition we call Reduced Energy Condition (REC).

We will show that it implies 
\begin{equation}\label{eq:intro-bound}
	m_\mathrm{KK}^2 \leqslant \alpha\max\left\{\sigma^2,\,\frac1{n-1}\left( | \Lambda | + \frac{\sigma^2}{D - 2} \right)\right\}+ \beta (m_D^{D-2}m_d^{2-d})^{2/n} 
\end{equation}
where $n=D-d$ is the internal dimension; $\alpha$ and $\beta$ only depend on $n$; $m_D$ and $m_d$ are the Planck masses in $D$ and $d$ dimensions respectively; and $\sigma\ge (D-2)|\dd A|$ is a bound on the gradient of the warping function $A$. The combination in the last term is natural; in the unwarped case, the parenthesis is the inverse volume. There is also a related bound, where $m_d$ is replaced by the diameter of the internal space $M_n$, i.e.~the largest distance between any two of its points. 
Even though our motivations come from the study of the problem of scale separation in string and M-theory, the bounds we derive can be of more general interest as they provide a precise relation among the different physical scales related in a compactification of any gravitational theory: the KK scale, the scale of the cosmological constant plus warping effects, and the reduced Planck mass (or diameter).

The REC is satisfied in $D=11$ supergravity, and in type II by everything except O-planes. It was indeed previously argued  \cite{gautason-schillo-vanriet-williams} that O-planes are necessary for scale separation; however, this required assuming $\sigma \ll 1$, and diagnosed separation based on the proxy $\int R^\mathrm{E}_6/\int R^\mathrm{E}_4$, the ratio of the integrated curvatures in Einstein frame. In many cases this proxy provides a good estimate, but it can easily fail, for example in presence of O-planes \cite[Sec.~2.2]{apruzzi-deluca-gnecchi-lomonaco-t}. In contrast, (\ref{eq:intro-bound}) is a mathematical statement on the mass of the lowest eigenvalue of a certain weighted Laplacian, which determines the masses of the spin-two fields \cite{bachas-estes,csaki-erlich-hollowood-shirman}. Indeed scale separation fails if a field of any spin has mass$^2$ $\sim |\Lambda|$.

The strategy involves showing a lower bound on the eigenvalues of the internal $R_{mn}- (D-2)\nabla_m \partial_n A$. This makes $M_n$ a so-called \emph{Bakry--\'Emery manifold}, on which several results are known, adapting some of which \cite{hassannezhad,setti-eigenvalues} we obtain (\ref{eq:intro-bound}) and the other bound we mentioned.

While these results constrain the KK scale, they still seem to allow scale separation, even without sources. In light of this, we revisit in Sec.~\ref{sec:sug} a phenomenon that first appeared in \cite[Sec.~7]{saracco-t}, where an O6-plane singularity can become smooth upon deformation by Romans mass $F_0\neq 0$. 
This happens when one adapts the supersymmetry equations to a situation where the O6 wraps an internal flat cycle. It is not to be understood as a quantum resolution of the O6, since it appears in supergravity. Rather, there is a topology change such that the O6 involution no longer has a fixed locus where an O6-plane could sit.

However, this smooth deformation was demonstrated in \cite[Sec.~7]{saracco-t} for a ``generic'' branch of the supersymmetry equations where $F_6\neq 0$; this had an exact NSNS three-form $H$, which obstructed promoting the local solution to a global one. It was realized more recently 
\cite{junghans-dgkt,marchesano-palti-quirant-t} that a special branch with $F_6=0$ does not have this problem. So we look again at the local behavior near an O6 on this special branch. Our result in Sec.~\ref{sec:sug} is that the O6 singularity this time gets modified by $F_0\neq 0$ to a different singularity, which can be formally interpreted as an O4 partially smeared along two directions.\footnote{It is worth recalling that in other contexts an O6 singularity is not modified at all by $F_0$, such as for AdS$_7$ solutions \cite{afrt,cremonesi-t,apruzzi-fazzi}.} Such singularities appeared elsewhere, for example in \cite{passias-prins-t}; they are still problematic, but they might be an intermediate step towards a better-defined solution. 

This result should be relevant in particular for the above-mentioned class of supersymmetric\footnote{As pointed out in \cite{polchinski-silverstein}, supersymmetry protects many BPS operator dimensions in a CFT, preventing them from getting large anomalous dimensions. The AdS dual of this statement is the presence of many fields with small masses, thus hindering scale separation. For example, scale separation was shown to be impossible for supersymmetric AdS$_7$ solutions \cite[Sec.~2.2]{apruzzi-deluca-gnecchi-lomonaco-t}. However, the solutions in \cite{dewolfe-giryavets-kachru-taylor} only have ${\mathcal N}=1$ supersymmetry in $d=4$, which does not protect any operators for the $d=3$ CFT duals.} solutions AdS$_4\times \mathrm{CY}_6$ of \cite{dewolfe-giryavets-kachru-taylor}. This was lifted to a IIA supergravity solution in \cite{acharya-benini-valandro}, but with smeared O6-planes. To find a more sensible lift with localized O6-planes, \cite[Sec.~6]{saracco-t} suggested to work in a limit where the curvature radii are large, while the string coupling $\ee^\phi$ is small almost everywhere. This idea was recently realized for the special branch with $F_6=0$, where an approximate large-volume solution was found \cite{junghans-dgkt,marchesano-palti-quirant-t}. 

To sum up, the results of Sec.~\ref{sec:be} in this paper constrain the masses of the KK modes in terms of certain geometric data and of the matter content of the theory, but leave open the possibility of scale separation in supergravity even without O-planes. This might have opened the possibility that the O6-planes in AdS$_4\times \mathrm{CY}_6$ solutions of \cite{dewolfe-giryavets-kachru-taylor} might have been smoothed out by $F_0$ as in \cite[Sec.~7]{saracco-t}; but the results in Sec.~\ref{sec:sug} seem to exclude this. Thus the status of scale-separated solutions in supergravity seems to rely on our understanding of O-plane singularities, as for de Sitter solutions.


\section{General bounds} 
\label{sec:be}

In this section, we will derive general bounds on the KK spectrum of a compactification. After reviewing the basics in Sec.~\ref{sub:kk},
we will describe the geometry of compactifications of higher dimensional gravitational theories down to $d$-dimensional vacua within the framework of \emph{Bakry--\'Emery manifolds}, defined in terms of a Riemannian manifold $M_g$ and a real function $f$. In Sec.~\ref{sec:BE} we will identify $M_g$ with the internal space of compactifications and $f$ with a multiple of the warping factor.
With this language we can describe many physical properties of the vacuum, such as the spectrum of spin-two excitations, in terms of natural objects in Bakry--\'Emery geometry.
We show in Sec.~\ref{sec:boundCurv} that the Bakry--\'Emery Ricci tensor can be bounded from below by using the equations of motion, if the stress energy tensor satisfies a Reduced Energy Condition (REC), which we analyze in Sec.~\ref{sec:stresses}. We then proceed in Sec.~\ref{sec:masses} to exploit some known mathematical results to bound the masses of the spin-two particles.

Albeit our motivations come from the study of compactifications of ten- and eleven-dimensional supergravity theories, which describe the low-energy limit of String Theories and M-theory respectively, the results we are going to present in this section are general and apply to any higher-dimensional gravitational theory that satisfies the REC. As we show in Sec.~\ref{sec:stresses} this is not restrictive, as most classical sources satisfy it.

\subsection{General expectations} 
\label{sub:kk}

We will consider a general compactification of a $D$-dimensional gravitational theory, with an Einstein--Hilbert (EH) term $m_D^{D-2} \int \dd^D x R_D$, down to $d$-dimensional vacua:\footnote{
	We use upper case Latin letters to denote $D$-dimensional indices, lower-case Greek letters for indices along the directions of the $d$-dimensional vacuum and lower-case Latin letters for
	indices in the $n$-dimensional internal space, with $n= D-d$.}
\begin{equation}\label{eq:metric}
	\dd s^2_D \equiv 
	\ee^{2 A} \dd  s^2_{d} + \dd s^2_n \equiv \ee^{2 A} (\dd  
	s^2_{d} + \bar{\dd s}^2_n)\,.
  \end{equation}
The warping function $A$ only varies over the $n$-dimensional internal space;  $\dd s^2_d$ is a maximally-symmetric space, with curvature normalized as $R^{(d)}_{\mu \nu} = \Lambda g^{(d)}_{\mu \nu}$. The ``barred'' internal metric $\bar g_{mn}= \ee^{-2A} g_{mn}$ will make our analysis easier later. The reduced action will have an EH term $m_d^{d-2}\int \dd^d x R_d$, with
\begin{equation}\label{eq:mdD}
	\frac{m_d^{d-2}}{m_D^{D-2}}=  \int \dd^n y \sqrt{g_n} \,\ee^{(d-2) A}= \int \dd^n y \sqrt{\bar g_n} \,\ee^{(D-2) A}\,
\end{equation}
relating the Planck masses in $D$ and $d$ dimensions.

Notice that in any warped compactification there is an ambiguity in defining the cosmological constant, since a change in its value could absorbed by a constant shift of the warping: in other words, (\ref{eq:metric}) is invariant under
\begin{equation}\label{eq:amb}
	A \to A+ A_0 \, ,\qquad	 g_{\mu \nu}\to \ee^{-2A_0}g_{\mu \nu}\, ,\qquad	 \bar{g}_{m n}\to \ee^{-2A_0}\bar{g}_{m n}\,.
\end{equation}
The mass scales we will consider do suffer from this ambiguity, but fortunately their ratios do not.

We are interested in the masses of the KK modes. These are obtained as eigenvalues of differential operators on $M_n$ similar to the Laplacian. Below we will use the lowest mass of the spin-two fields as a definition of $m_\mathrm{KK}$; notice that making $m_\mathrm{KK}/m_{\Lambda}$ large, $m_\mathrm{\Lambda}\equiv \sqrt{|\Lambda|}$ is a necessary condition for scale separation to hold. 

Many AdS vacua only have one length scale $r$; obviously then $m_\Lambda$, $m_d$, $m_\mathrm{KK}$ are all of order $1/r$ and there is no scale separation. More generally, if one assumes that the internal Ricci tensor has a positive bound from below, $R_{mn}\ge r^{-2} g_{mn}$ for some $r$, famous results \cite{lichnerowicz-bound,li-yau} then imply bounds on the Laplacian. If moreover one also assumes that the gradient $\partial_m A$ of the warping is negligible, then the equations of motion imply \cite{gautason-schillo-vanriet-williams} $r\sim R_\mathrm{AdS}$, the AdS length scale. But in general the Ricci tensor is not positive; it cannot be negative-definite either \cite{douglas-kallosh}, but the indefinite case is allowed. One can generate many such examples via the map
\begin{equation}\label{eq:hyp-map}
	\mathrm{AdS}_d\times M_{D-d} \to \mathrm{AdS}_{d'}\times H_{d-d'}\times M_{D-d}
\end{equation}
where $H_{d-d'}$, $d-d'>1$ is a compact space satisfying $R_{mn}= \Lambda g_{mn}$; for example for $d-d'=2$ this can be a Riemann surface with genus $g>2$ (since $\Lambda<0$). Indeed the equations of motion are completely unchanged under (\ref{eq:hyp-map});\footnote{This is not to be confused with the more sophisticated map induced holographically by compactifying a CFT$_{d-1}$, which changes the metric on $M_{D-d-d'}$ and fibers it over $H_{d-d'}$, and preserves supersymmetry \cite{maldacena-nunez}. (\ref{eq:hyp-map}) was used for example in searching dS$_4$ solutions with O6-planes \cite{cordova-deluca-t-dso6}.} even if $M_{D-d}$ is positively curved in the original solution, this map produces an example where $R_{mn}$ also has negative eigenvalues.

There do also exist results on the Laplacian operator when the Ricci tensor has a negative bound from below \cite{cheng-bound}. However, when $A$ is not constant, it is difficult to obtain such a bound from the equations of motion; moreover, we will see below that $A$ also enters in the differential operators whose eigenvalues give the KK masses. 

Fortunately, the two issues are related; the KK operator is natural in a mathematical framework called Bakry--\'Emery geometry. This allows to derive eigenvalue results, if a bound exists for the combination $R_{mn}- (D-2) \nabla_m \nabla_n A$ of the Ricci tensor and warping, which is precisely of the type provided by gravity compactifications.


\subsection{Bakry--\'Emery geometry and compactifications}\label{sec:BE}

A Bakry--\'Emery manifold is just a pair of a Riemannian manifold and a real scalar function on it; the latter will be for us proportional to the warping function.

The spectrum of excitations around any compactification \eqref{eq:metric} includes a tower of spin-two particles defined as perturbations $\delta g_{\mu \nu}^{(d)}$ of the maximally symmetric metric $g^{(d)}_{\mu \nu}$.
Well-known work \cite{csaki-erlich-hollowood-shirman,bachas-estes} shows that, unlike for lower spins, the spin-two tower always decouples and can be studied independently, without needing to first diagonalize the full spectrum.
Moreover, the operator whose eigenvalues give the masses of these spin-two particles has a universal form depending only on the internal metric and on the warping factor.

This property has allowed a relatively easy computation of the spin-two spectrum of many warped AdS vacua where a full diagonalization of the spectrum is currently out of reach; see for example \cite{klebanov-pufu-rocha,ahn-woo,richard-terrisse-tsimpis, passias-t, pang-rong-varela, gutperle-uhlemann-spin2, passias-richmond, apruzzi-deluca-gnecchi-lomonaco-t}.\footnote{For large supersymmetry, recently a different approach has been developed based on exceptional geometry, where the full spectrum can be computed for all vacua based on a given internal manifold \cite{malek-samtleben-kk,malek-nicolai-samtleben}.} Recent studies of how extra dimensions would influence gravity waves also include \cite{andriot-lucenagomez,andriot-tsimpis,andriot-marconnet-tsimpis}. The effect of warping on KK reductions was also studied earlier in \cite{giddings-maharana,douglas-shiu-torroba-underwood}, and its connection with the problem of scale separation in \cite{tsimpis-scale}.

There are many equivalent ways to rewrite the spin-two operator \cite{bachas-estes}.
A natural form from our current perspective is
\begin{equation}\label{eq:BELaplacian}
  \Delta_f (\psi) \equiv - \frac{1}{\sqrt{\bar{g}}} \ee^{- f} \partial_m
  \left( \sqrt{\bar{g_{}}}  \bar{g}^{m n} \ee^f \partial_n \psi \right) = \Delta \psi-\bar{\nabla} f \cdot \dd \psi\;,
 \qquad f = (D - 2) A \;.
\end{equation}
This is called \emph{Bakry--\'Emery Laplacian} \cite{bakry-emery};\footnote{When acting on general $k$-forms, this operator can also be written as $\Delta_f = \dd_f^\dagger \dd + \dd \dd^\dagger_f$ with
 $\dd_f^\dagger \equiv \ee^{-f} \dd^{\dagger} \ee^f$.
This form makes it closely related to the operator $\dd_f^\dagger \dd_f + \dd_f \dd^\dagger_f $, with $\dd_f\equiv \ee^{f}\dd_f \ee^{-f}$,  which has been used by Witten to prove the Morse inequalities \cite{witten-morse}.} the spin-two modes $h_{\mu \nu}^k$ and their masses $m^2_k$ are given by 
\begin{equation}
	\Delta_f \psi_k = m_k^2 \psi_k\,.
\end{equation}

Even without working out the full operator as in \cite{bachas-estes}, the weight $\ee^{(D-2)A}$ can be understood by rewriting the $D$-dimensional Einstein-Hilbert action in terms of the $D$-dimensional
unwarped metric $\bar{g}_D \equiv \ee^{-2A} g_D$:
\begin{equation}\label{eq:unwarpedEH}
	S_\text{EH} = m_D^{D-2}	\int \sqrt{-\bar{g}_D}\ee^{(D-2)A} (\bar{R}_D+\ldots)\;,
\end{equation} where the dots refer to terms proportional to derivatives of $A$.
Now $\bar{g}_D$ is a pure (unwarped) product, and the spin 2 fields are simply obtained by varying \eqref{eq:unwarpedEH} at second order with respect to the $d$-dimensional part of $\bar{g}_D$.
This shows that the internal differential operators are naturally weighted with the factor $\ee^{(D-2)A}$.
Although lower spins do not decouple and one needs to diagonalize their action before being able to read their mass operators, a similar argument suggests they can be related to Bakry--\'Emery Laplacians, possibly with a different weight.
For example, a $k$-flux term $S_{F_k} \equiv -\frac{m_D^{D-2}}{2} \int \sqrt{-g_D} F_k^2$ will be accompanied by an $\ee^{(D-2 k)A} $ factor coming from the inverse metrics appearing in $F_k^2$ .
Aside from the zero-mode, the bottom of the spectrum of $\Delta_{c f}$ can be seen to be lower than the one of $\Delta_f$ if $c<1$ by rewriting the eigenvalue equation as a Schr\"odinger problem, as in \cite[Section 3]{hassannezhad}.

As it is the case in usual Riemannian geometry, the spectrum of the Laplacian is controlled by the curvature.
In Bakry--\'Emery geometry a natural notion of curvature is the \emph{Bakry--\'Emery Ricci curvature}, defined as $\text{Ricci}_f\equiv \text{Ricci} - \text{Hess}(f)$, or in index notation:\footnote{This tensor is sometimes called $\infty$-Bakry--\'Emery in the literature. The name comes from viewing it as a special case of the $N$-Bakry--\'Emery-Ricci tensor 
\mbox{$R_{m n}-\nabla_m \nabla_n f-\frac{1}{N} \nabla_m f \nabla_n f$}.
As we are going to see in Sec.~\ref{sec:boundCurv}, the equations of motion naturally bound the $\infty$-Bakry--\'Emery-Ricci curvature, and we will thus avoid this notation and simply call it Bakry--\'Emery Ricci tensor.}
\begin{equation}\label{eq:RiccifDef}
	R_{m n}-\nabla_m \nabla_n f\,.
\end{equation}

Many standard geometric results that apply in the pure Riemannian case ($f=0$)
often carry over to the Bakry--\'Emery case, by replacing the Ricci tensor with the Bakry--\'Emery Ricci tensor \eqref{eq:RiccifDef} and requiring a bound on $(\nabla f)^2$.
We are going to present some of these results in Sec.~\ref{sec:masses}, where we will use them to show how a lower bound on the Bakry--\'Emery curvature \eqref{eq:RiccifBound} translates into an upper bound on all the eigenvalues of the operator \eqref{eq:BELaplacian}.

We will now bound Ricci$_f$ in terms of an energy condition on the stress-energy tensor of the theory, which we will check in \ref{sec:stresses} for various matter sources.

\subsection{Bounding the curvature}\label{sec:boundCurv}
We start from the $D$-dimensional Einstein equations written as\footnote{
We normalize the Einstein-Hilbert term in the action as $S_{\text{EH}} = \frac{1}{\kappa^2}\int\sqrt{-g_D}R_D$ with $\kappa^2 = (2\pi)^{D-3}\ell_D^{D-2}$, where $\ell_D$ is the $D$-dimensional Planck length.
}

\begin{equation}\label{eq:RMN}
  R_{M   N} = \frac{1}{2}\kappa^2 \left( T_{M   N} - g  _{M  
  N} \frac{T}{D-2}  \right) \equiv \hat{T}_{M N}
\end{equation}
where $T_{M N} \equiv -\frac{2}{\sqrt{-g}} \frac{\delta S_{\text{mat}}}{\delta g^{M N}}$ is the stress-energy tensor of the $D$-dimensional theory.

To specialize \eqref{eq:RMN} to metrics of the form \eqref{eq:metric}, we first compute the internal and external components of the Ricci tensor:
\begin{align}
	\label{eq:ExtEin}
  R_{\mu \nu}^{(D)} & =  g_{\mu \nu}^{(d)} (\Lambda - \bar{\nabla}^p \bar{\nabla}_p A -
  (D - 2) \bar{\nabla}_p A \bar{\nabla}^p A)  \\
  \label{eq:IntRmn} R_{m n}^{(D)} & =  \bar{R}_{m n}^{} - (D - 2) \bar{\nabla}_m \bar{\nabla}_n A + (D - 2)
  \bar{\nabla}_m A \bar{\nabla}_n A + \\
  & -  \bar{g}_{m n} (\bar{\nabla}^p \bar{\nabla}_p A + (D - 2) \bar{\nabla}_p A \bar{\nabla}^p A) .
  \nonumber
\end{align}
Let us first focus on the external equation obtained from \eqref{eq:ExtEin}.
Since the external space is maximally-symmetric, if the stress-energy tensor along these directions also respects this symmetry, we lose no information by taking its trace, obtaining the scalar equation
\begin{equation}\label{eq:extEq}
  \Lambda - \bar{\nabla}^2 A - (D - 2) (\bar{\nabla} A)^2 = \frac{1}{d} \hat{T}^{(d)}\,,
\end{equation}
where $\hat{T}^{(d)} \equiv g^{(d)\, \mu \nu}\hat{T}_{\mu \nu}$;
equivalently,
\begin{equation}
  \Lambda - \frac{1}{D - 2} \ee^{- (D - 2) A} \bar{\nabla}^2 (\ee^{(D - 2)
  A}) = \frac{1}{d} \hat{T}^{(d)}\,. \label{eq:MN2}
\end{equation}
If the internal space $M_n$ is smooth and compact, integrating this gives
\begin{equation}
	\Lambda \int_{M_n} \dd^n y \sqrt{\bar g} \ee^{(D - 2) A}  =  \int_{M_n} \dd^n y \sqrt{\bar g} \ee^{(D - 2) A} \hat{T}^{(d)}\;.
\end{equation}
This equation is often used to derive no-go theorem for dS compactifications of supergravity theories \cite{gibbons-nogo,maldacena-nunez, dewit-smit-haridass}, 
since for most of the classical sources of eleven and ten-dimensional supergravity theories the right hand side is non-positive.
This excludes O-planes, which also violate the assumption of a smooth internal space without boundaries; and the Romans mass $F_0$, which was however excluded in \cite{maldacena-nunez} with a separate analysis.

The internal equations (\ref{eq:IntRmn}) cannot be reduced to a scalar equation without loss of information. Some constraints can be obtained by taking its trace \cite{douglas-kallosh, gautason-schillo-vanriet-williams, petrini-solard-vanriet}, but the Ricci scalar alone often gives too weak a restriction on the geometry for most geometrical purposes.

Keeping then all the internal directions, the internal equation reads
\begin{equation}\label{eq:IntEin}
  \bar{R}_{m n} - (D - 2) \bar{\nabla}_m \bar{\nabla}_n A = (D - 2) (  -
  \bar{\nabla}_m A \bar{\nabla}_n A + \bar{g}_{m n}  (\bar{\nabla} A)^2) +
  \bar{g}_{m n} \bar{\nabla}^2 A + \hat{T}^{(D)}_{m n}\;.
\end{equation}
The left hand side is of the form \eqref{eq:RiccifDef}, but in order to put a bound on it we need to get rid of
the $\nabla^2A$ term on the right hand side, which does not have a definite sign.
We achieve this by using the external equation \eqref{eq:extEq}:
\begin{equation}\label{eq:pieces}
	\begin{split}
		\bar{R}_{m n} - (D - 2) \bar{\nabla}_m \bar{\nabla}_n A &=\Lambda
		  \bar{g}_{m n}  +\\
		&- (D - 2) (\bar{\nabla}_m A
		\bar{\nabla}_n A) + \\
		&+ \hat{T}^{(D)}_{m n} -
		\bar{g}_{m n} \frac{1}{d} \hat{T}^{(d)}		\;.
	\end{split}
\end{equation}
We now analyze the various terms on the right hand side. 
The first line is responsible for relating the scales of the internal and external curvature.
The second line is a negative-definite quadratic form, whose single non-zero eigenvalue is the square of the gradient of the warp factor. This will play an important role, and we will call $\frac{\sigma^2}{(D-2)^2}\equiv \mathrm{sup}_{M_n} \bar g^{mn} \partial_m A \partial_n A$. In other words:
\begin{equation}\label{eq:bounddA}
	(D-2)|\bar{\nabla} A|  \leqslant \sigma .
\end{equation}
Finally, in the third line of \eqref{eq:pieces} we have the combination of stress-energy tensors
\begin{equation}\label{eq:comb1}
	\hat{T}^{(D)}_{m n} - \bar{g}_{m n} \frac{1}{d} \hat{T}^{(d)} =
	\frac{1}{2 }\kappa^2 \left(T^{(D)}_{m n} - \bar{g}_{m n} \frac{1}{d} T^{(d)}\right) \;.
\end{equation}
As we are going to see in the next section, this quantity is non-negative for a broad class of matter fields; this includes those of $d=10$ and $d=11$ supergravity, with the only exception of O-planes. Our assumptions are also invalidated by string or M-theory corrections, where the kinetic term is not even of EH type. In this sense, our results will be related to those for dS compactifications \cite{gibbons-nogo,dewit-smit-haridass,maldacena-nunez}. However, requiring that \eqref{eq:comb1} is non-negative is inequivalent to the Strong Energy Condition, which is violated by a $D$-dimensional cosmological constant (such as the Romans mass in massive IIA supergravity) while \eqref{eq:comb1} is not.
Since the combination \eqref{eq:comb1} only makes sense when reducing the higher-dimensional theory to lower dimensional vacua, we will call the corresponding condition \emph{Reduced Energy Condition} (REC):
\begin{equation}\label{eq:Econd}
	\text{REC:} \qquad  T^{(D)}_{m n} - \bar{g}_{m n} \frac{1}{d} T^{(d)}\geqslant 0 \;.
  \end{equation}
Since, as we mentioned, effects that violate (\ref{eq:Econd}) are also needed to obtain dS compactifications, from now on we will take $\Lambda<0$.

Summing up, if the Reduced Energy Condition \eqref{eq:Econd} holds, there is a bound on the Bakry--\'Emery Ricci tensor:
\begin{equation}
	\text{Ricci}_f \geqslant -\left( | \Lambda | + \frac{\sigma^2}{D - 2}
	  \right)\,,
\end{equation}
with $f=(D-2)A$. More explicitly, recalling \eqref{eq:RiccifDef}:
\begin{equation}\label{eq:RiccifBound}
  \bar{R}_{m n} - (D - 2) \bar{\nabla}_m \bar{\nabla}_n A \geqslant - \left( | \Lambda | + \frac{\sigma^2}{D - 2} 
  \right)\bar g_{mn}\,.
\end{equation}
Of course this bound is only useful if $\sigma$ defined in \eqref{eq:bounddA} is finite. We will use this information in Sec.~\ref{sec:masses}, but let us first check the REC for various sources.

\subsection{Stress-energy contributions}\label{sec:stresses}

In this section, we verify that the Reduced Energy Condition \eqref{eq:Econd} is satisfied for a broad class of contributions to the stress-energy tensor, including those in $d=10$ and $d=11$ supergravity (with the exception of O-planes).

We start with form field-strengths, with an action of the form
\begin{equation}\label{eq:SFk}
	S_{F_k} = - \frac{1}{2 \kappa^2} \int \dd^D x \sqrt{- g} \,\ee^{\lambda \phi} F_k^2\;,
\end{equation}
where $\ee^{\lambda \phi}$ is a possible scalar weight. This includes the action for NSNS and RR fluxes in $d=10$ supergravity, and the four-form flux in $d=11$. The square of a form is defined as
\begin{equation}
	\begin{array}{lll}
	F_k^2 & \equiv & \frac{1}{k!} (F_k)_{M_1 \ldots M_k} (F_k)_{P_1 \ldots P_k}
	g^{M_1 P_1} \dots g^{M_k P_k} \;.
  \end{array}
\end{equation}
More generally we define the inner product $F \cdot \tilde F\equiv \frac1{k!} F_{M_1\ldots M_k} \tilde F^{M_1\ldots M_k}$; in particular, if $(F_k)_M \equiv \iota_M F_k$, we have
\begin{equation}
	(F_k)_M \cdot (F_k)_N = \frac{1}{(k - 1) !} (F_k)_{M P_1 \ldots P_{k - 1}} (F_k)_N^{\quad P_1 \ldots
	P_{k - 1}}\,.
\end{equation}
This tensor vanishes for a 0-form, since $\iota_M F_0=0$.

In order not to break maximal symmetry of the vacuum, a $k$-form can always be decomposed as
\begin{equation}
  F_k \equiv f_\mathrm{e} \wedge \text{vol}_d + f_\mathrm{i}
\end{equation}
where $f_\mathrm{e}$ and $f_\mathrm{i}$ are differential forms defined purely on the internal $n$-dimensional space.
Notice that by construction $f_\mathrm{e} \neq 0$ only if $k\geqslant d$. (In type II supergravity, $f_\mathrm{e}$ and $f_\mathrm{i}$ are Hodge dual to each other in $M_n$.)

With these definitions, the contractions that appear in the stress-energy tensor read
\begin{eqnarray}\label{eq:Fkcontractions}
  F_k^2 & = & f_\mathrm{i}^2 - \ee^{- d A} f_\mathrm{e}^2 \\
  (F_k)_{\mu} \cdot (F_k)_{\nu} & = & - g_{\mu \nu} \ee^{- d A} f_\mathrm{e}^2  \\
  (F_k)_m \cdot (F_k)_n & = & (f_\mathrm{i})_m \cdot (f_\mathrm{i})_n - \ee^{- d A} (f_\mathrm{e})_m \cdot
  (f_\mathrm{e})_n  \;.
\end{eqnarray} 
Varying \eqref{eq:SFk} we compute the stress-energy tensor for a $k$-dimensional flux $F_k$.
Using \eqref{eq:Fkcontractions} the combination \eqref{eq:comb1} reads
\begin{equation}\label{eq:combStressF}
	 \kappa^2 \left( T_{m n} - \frac{1}{d} \bar{g}_{m n} T^{(d)} \right)_{F_k} =
	 \ee^{- 2 A (k - 1)}\ee^{\lambda \phi}\left[ (\bar{f}_i)_m \cdot (\bar{f}_i)_n - 
	 (\bar{f}_\mathrm{e})_m \cdot (\bar{f}_\mathrm{e})_n+\bar{g}_{m n}  \bar{f}_\mathrm{e}^2\right]\;,
\end{equation}
where a bar over $f$ reminds us that the contractions in the products are taken with the internal $\bar{g}$.
Notice from \eqref{eq:combStressF} that a cosmological constant term (such as the Romans mass $F_0$ in type IIA) does not contribute to the REC.
Similarly, the contribution to the dilaton potential that is present in string theories in non-critical dimension does not contribute to the REC.
That the quantity \eqref{eq:combStressF} is non-negative can be seen as follows.
For any internal form $f$, an explicit computation reveals that $(\star f)^2 = f^2$ and $-\bar{f}_m\cdot \bar{f}_n+ \frac12 \bar{g}_{m n} \bar{f}^2 =  (\star \bar{f})_m\cdot (\star \bar{f})_n- \frac12 \bar{g}_{m n} (\star \bar{f})^2$, where
$\star$ is computed with the internal metric $\bar{g}$.
Applying these relations to the $f_\mathrm{e}$ term in \eqref{eq:combStressF} gives
\begin{equation}\label{eq:combStressF2}
	\kappa^2 \left( T_{m n} - \frac{1}{d} \bar{g}_{m n} T^{(d)} \right)_{F_k} =
	\ee^{- 2 A (k - 1)}\ee^{\lambda \phi}\left[ (\bar{f}_\mathrm{i})_m \cdot (\bar{f}_\mathrm{i})_n +
	(\star\bar{f}_\mathrm{e})_m \cdot (\star\bar{f}_\mathrm{e})_n \right] \geqslant 0\;.
\end{equation}

We now turn our attention to a canonically normalized scalar field $\phi$, such as the dilaton in string theory in Einstein frame:
\begin{equation}\label{eq:Sphi}
	S_{\phi} = - \frac{1}{2 \kappa^2} \int\dd^D x \sqrt{- g} (\nabla \phi)^2	\;.
\end{equation}
Formally, \eqref{eq:Sphi} has the same structure as \eqref{eq:SFk} for $k=1$, upon the identification $F_1 =\nabla\phi$.
Since in order to do not break the maximal $d$-dimensional symmetry $\phi$ is only allowed to depend on the internal coordinates,
we obtain the same result as in  \eqref{eq:combStressF}, specialized to a purely internal $F_1$:
\begin{equation}
	\kappa^2 \left( T_{m n} - \frac{1}{d} \bar{g}_{m n} T^{(d)} \right)_\phi = 
	\partial_m \phi \partial_n \phi
	\geqslant 0.
\end{equation}

Finally, we consider localized sources, with an action of the form
\begin{equation}
S_p = 
- \tau_p \int_{\Sigma_{p + 1}}\dd^{p+1} \sigma \ee^{\tilde\lambda \phi}
 \sqrt{- g |_{\Sigma p + 1}}\;.
\end{equation}
This includes D-branes in string theory and M-branes in M-theory, again with $\phi$ playing the role of the dilaton; the parameter $\tilde\lambda=(p-3)/4$ in Einstein frame.  Separating the stress-energy tensor in directions parallel and transverse to the worldvolume,
\begin{equation}
	\begin{array}{llll}
		T_{i j} & = & - \frac{1}{2} \,\tau_p\, \rho \ee^{\tilde{\lambda} \phi} g_{i j} 
		(g_{\perp})^{- 1 / 2} &  i, j \quad \text{parallel}\;,\\
		T_{i j} & = & 0 & i, j\quad \text{transverse}\;.
	\end{array}
\end{equation}
We introduced a density function $\rho$ in the internal space, which reduces to a Dirac $\delta$ for completely localized sources.

In order not to break maximal $d$-dimensional symmetry, we assume the source is parallel to the all the vacuum directions.
Evaluating the combination \eqref{eq:comb1} we get
\begin{equation}\label{eq:combLoc}
	\begin{array}{llll}
		T^{(D)}_{m n} - \frac{1}{d} \bar{g}_{m n} T^{(d)} & = & 0& m, n \quad \text{parallel}\;,\\
		T^{(D)}_{m n} - \frac{1}{d} \bar{g}_{m n} T^{(d)} & = & \tau_p \,\rho\, \ee^{\tilde\lambda \phi} \ee^{- (7 - p) A} \bar{g}_{m n} 
		(\bar{g}_{\perp})^{- 1 / 2}& m, n \quad \text{transverse}\;.
	\end{array}
\end{equation}
From \eqref{eq:combLoc} we see that for positive tension, $\tau_p > 0$, the Reduced Energy Condition \eqref{eq:Econd} is satisfied, while it is violated by negative tension objects, such as O$p$-planes in string theory.

As we mentioned earlier, string and M-theory corrections to $d=10$ and $d=11$ --- and quantum gravity effects more generally --- are not included in our discussion.
Indeed, it can be easily checked that Casimir energies can violate the REC \eqref{eq:Econd}.
As an example, inclusion of Casimir energies in the Standard Model lead to AdS$_3\times S^1$ vacua with scale separation \cite{arkanihamed-dubovsky-nicolis-villadoro}.
Similarly, by including Casimir energies in the internal space, one can construct scale-separated AdS vacua in String and M-theory with a bounded warping factor \cite{toappear-casimir}.

\subsection{Physical implications}\label{sec:masses}

A lower bound on the Bakry--\'Emery curvature has various geometrical consequences. In the context of compactifications these directly translate into physical properties of the lower-dimensional vacuum. In particular we will now see that the bound \eqref{eq:RiccifBound} implies some inequalities on spectrum of spin-two fields, whose masses are given by the eigenvalues of the Bakry--\'Emery Laplacian (\ref{eq:BELaplacian}).

In particular, these are interesting to us because for scale separation the masses of \emph{all} the Kaluza-Klein modes have to be much bigger than the mass scale of the $d$-dimensional cosmological constant:
\begin{equation}
	\text{scale-separated vacuum:}\qquad \frac{m_k^2}{|\Lambda|} \gg 1\,.
\end{equation} 

A first spectrum inequality we find uses \cite[Th.~A.1]{hassannezhad}:
\begin{theorem}\label{thm:boundVol}
	  Any compactification of a $D$-dimensional gravitational theory whose stress energy tensor satisfies the Reduced Energy Condition \eqref{eq:Econd} and with warping bounded by $(D-2)|\bar\nabla A| \leqslant \sigma$ has a spectrum of spin-two excitations whose masses are bounded by
	\begin{equation}\label{eq:asma}
		m_k^2 \leqslant \alpha(n)\max\left\{\sigma^2,\,\frac1{n-1}\left( | \Lambda | + \frac{\sigma^2}{D - 2} \right)\right\} 
		+ \beta(n)
		\left(k\, 
		\frac{\mathrm{sup}(\ee^{(D-2)A})}{\int \dd^n y \sqrt{\bar g_n}\,\ee^{(D-2)A}}
		\right)^{2/n}\,.
	\end{equation}
\end{theorem}
The behavior $k^{2/n}$ is the one expected from the Weyl asymptotic law for the eigenvalues of the Laplacian. Recalling (\ref{eq:mdD}), the integral in the second term is $m_d^{d-2}/m_D^{D-2}$, where recall $m_D$ and $m_d$ are the Planck masses for the gravity theory before and after compactification. This reproduces the bound in (\ref{eq:intro-bound}) for the overall KK scale. It is also immediate to verify that $m_k^2\to \ee^{2A_0} m_k^2$ under (\ref{eq:amb}), so the bound on $m_k^2/m^2_\mathrm{\Lambda} = m_k^2/ |\Lambda|$ is unambiguous, as promised.

The constants $\alpha(n)$ and $\beta(n)$ can be estimated using their definitions in \cite[App.~A]{hassannezhad}:\footnote{We thank A.~Hassannezhad for her patient explanation of her work.}
\begin{equation}
	\alpha(n)< 16 \cdot 2^n \ee^{\frac{n+1}2} \, ,\qquad \beta(n)< 2^{4+n +\frac2n}  \ee^{\frac{n+5}2} \pi (n\, \Gamma(n/2))^{-\frac2n}\,. 
\end{equation}
In particular for $n=6$, $7$ they are bounded by numbers of order $10^4$ and $10^5$ respectively.

Focusing on string theory, recall that we ran our general argument in the previous subsections with an EH kinetic term, so $A=A_\mathrm{E}$ is the warp function in Einstein frame; the integral in (\ref{eq:asma}) and the gradient norm $|\bar \nabla A|$ are computed in the barred metric defined in (\ref{eq:metric}).
Even though the mathematical theorems behind Th.~\ref{thm:boundVol} are formulated in the smooth setting, we can check whether these quantities are finite and if the statements make sense for some physical singularities, such as D-branes in string theory. Near a D$p$-brane, the ten-dimensional Einstein frame metric asymptotes to 
\begin{equation}\label{eq:dsbrane}
	\dd s^2_{10} \sim H^{\frac{p-7}{8}}\left( \dd x^2_{p+1}+ H(\dd r^2+r^2 \dd s^2_{\mathbb{S}^{8-p}}) \right)\qquad \text{for}\; r\to 0\;,
\end{equation} 
where $H$ is a harmonic function in the transverse directions (prior to back-reaction): $H\sim \left(\frac{r}{r_0}\right)^{p-7}$ for $ p\neq 7$ and $H\sim -\log{(r/r_0)}$ for $p = 7$. We can then use \eqref{eq:dsbrane} to extract the barred metric \eqref{eq:metric} as well as the asymptotic weight function $\ee^f \sim H ^{\frac{p-7}{2}}$. An explicit computation then reveals that for $p\leqslant 5$ (and trivially for $p=7$), $\sigma^2$ stays finite approaching the singularity, and thus the general bound \eqref{eq:RiccifBound} on Ricci$_f$ is still meaningful. For this reason, we expect that the above results also hold for compactifications with such branes. For $p = 6$, instead, $\sigma^2$ diverges, and as currently stated the theorem becomes empty. An explicit computation shows that Ricci$_f$ is still bounded from below approaching a D6 brane, and thus an adapted formulation of the statement might still hold, but at the moment we do not have a candidate version.
In any case, consider that when singularities are present the eigenvalue problem needs to be formulated with care, as the spectrum could also turn out to be a continuum (for example, for $p\leqslant 5$, a singularity of the form \eqref{eq:dsbrane} makes the space non-compact, being at infinite distance in the barred metric.) 
We refer the interested reader to \cite{deluca-deponti-mondino-t}, where we discuss a more general mathematical framework in which singularities relevant for compactifications with brane sources are naturally included. This allows to obtain rigorous eigenvalue bounds even in the non-smooth setting. Also, notice that we have excluded O$p$-planes from the present discussion (recall that those violate the REC) but for $p\leqslant 5$ Ric$_f$  is bounded from below and they might be included in this framework \cite{deluca-deponti-mondino-t2}.

A second bound in the literature makes reference to the diameter of $M_n$, the largest possible distance between two points. This uses \cite[Cor.~4.4]{setti-eigenvalues}:
\begin{theorem}\label{thm:boundDiam}
	Any compactification of a $D$-dimensional gravitational theory whose stress energy tensor satisfies the Reduced Energy Condition \eqref{eq:Econd} and with warping satisfying $(D-2)|\bar\nabla A| \leqslant \sigma$  has a spectrum of spin-two excitations whose masses are bounded by
\begin{equation}\label{eq:setti}
	m_k^2\leqslant n \left( | \Lambda | + \frac{D - 1}{D - 2} \sigma^2 \right) + \gamma(n) \frac{k^2}{\mathrm{diam}(M_n)^2}\,.
\end{equation}
\end{theorem}
Once again the diameter should be computed with the metric $\bar{g}$. For the constant $\gamma(n)$ in the second term, \cite[Cor.~4.4]{setti-eigenvalues} quotes a limit involving an eigenvalue problem with $A=0$ on a ball in hyperbolic space, which from \cite[Cor.~2.3]{cheng-bound} is taken to be $4(1+2^m)^2\pi^2$ for $n=2(m+1)$ and $4(1+\pi^2) (1+2^{2m})^2$ for $n=2m+3$.

To our knowledge, the bounds (\ref{eq:asma}), (\ref{eq:setti}) are the first rigorous results valid in general for the KK scale in gravity compactifications. One might now hope to use them to show even sharper results; for example, it is natural to conjecture \cite{gautason-schillo-vanriet-williams} that scale separation is impossible in type II or $d=11$ supergravity without O-planes. 

Unfortunately this is unlikely with these bounds alone. To see why, consider for example the second term in (\ref{eq:asma}):
for $A=0$, it is $\mathrm{Vol}(M_n)^{-2/n}$. If we knew that this is of order $1/R_\mathrm{AdS}$, we would be done. But even for positive-definite Ricci tensor, it is easy to come up with sequences of vacua that have arbitrarily small volume by taking orbifolds; for example AdS$_4\times S^7/\mathbb{Z}_p$ have $\mathrm{Vol}(M_n)\sim R^7_\mathrm{AdS}/p$ (see also \cite{buratti-calderon-mininno-uranga}). In this case the bound (\ref{eq:asma}) gets increasingly useless with larger $p$. 

The second bound (\ref{eq:setti}) looks more promising in this respect, since the diameter is unchanged by taking quotients. There also exists a Bakry--\'Emery version of Myers's theorem \cite{wei-wylie} which puts a bound on the diameter; but just like the original with $f=0$, it only works when there is a lower \emph{positive} bound on the eigenvalues of $R_{mn} - \nabla_m \nabla_n f$. Our lower bound (\ref{eq:RiccifBound}) is negative, and a bound on the diameter cannot exist in this case, as the case of a torus with $f=0$ readily demonstrates. It is conceivable that a diameter bound might be proven by using (\ref{eq:RiccifBound}) in conjunction with other finer properties of $R_{mn}$, perhaps going back to the Raychaudhuri equation; unfortunately we do not see a way to prove this at this time.

Notice also that a \emph{lower} bound also exists in terms of the diameter \cite[Th.~3]{charalambous-lu-rowlett}, stating that the lowest spin-two mass satisfies the following
\begin{theorem}\label{thm:boundDiam-lower}
	Any compactification of a $D$-dimensional gravitational theory whose stress energy tensor satisfies the Reduced Energy Condition \eqref{eq:Econd} and with warping satisfying $(D-2)|\bar\nabla A| \leqslant \sigma$  has the first non-trivial spin 2 mass bounded by
\begin{equation}\label{eq:lowerBound}
	m_1^2 \geqslant \frac{\pi^2}{\mathrm{diam}^2}\exp\left(-c(n)\, \mathrm{diam} \sqrt{| \Lambda | + \frac{\sigma^2}{D - 2}}\right)\,,
\end{equation}
where $c(n)$ again only depends on the internal space dimension. 
\end{theorem}
The positive constant $c(n)$ can be estimated from the proof of \cite[Th.~3]{charalambous-lu-rowlett}. For $n\geqslant 3$, we obtain $c(n) = \frac{n-1}{2}\left(1+\frac{4}{\pi}\sqrt{\frac{2}{n-1}}\right)$. So we know that achieving small diameter would indeed prove scale separation for the spin-two fields.

The cautious words below equation \eqref{eq:dsbrane} on possible extensions to spaces that include brane-singularities also hold for Th.~\ref{thm:boundDiam} and \ref{thm:boundDiam-lower}. In particular, the fact that D$p$-brane sources for $p\leqslant 5$ are at infinite distance in the barred metric makes the diameter unbounded. This suggests, for example, a version of \eqref{eq:setti} without the second term, and indeed for smooth non-compact Bakry--\'Emery manifold such a theorem exists \cite[Thm.~1.1]{wu2013upper} and can be readily applied to this case.
We refer again to future work \cite{deluca-deponti-mondino-t2} where we will try to address the questions raised by physical singularities more systematically. 

To summarize, in this section we have shown that the masses of the spin 2 KK modes are bounded by the cosmological constant, the gradients of the warping function and the reduced Planck mass (Th.~\ref{thm:boundVol}) or the internal diameter (Th.~\ref{thm:boundDiam} and \ref{thm:boundDiam-lower}).
These results apply to any theory whose matter content satisfies the REC \eqref{eq:Econd}, and are currently formulated for smooth backgrounds, although we expect that physical singularities can be included as well \cite{deluca-deponti-mondino-t,deluca-deponti-mondino-t2}. In particular, this is true for any classical source in ten- and eleven-dimensional supergravities, except for O-planes.
This can be useful to constrain separation of scales, but more generally it provides precise relations among the different physical scales.


\section{O6-planes and Romans mass} 
\label{sec:sug}

The general bounds in the previous section do not exclude the possibility that scale separation might exist even without singularities. For this reason, in this section we will study the behavior of O6-planes in presence of Romans mass $F_0\neq 0$, completing \cite{saracco-t}. 

\subsection{Review} 
\label{sub:rev}

Let us first give a review of the relevant literature, to motivate our study.

The combination of O6-planes and Romans mass appears to be promising in achieving scale separation (as well as de Sitter solutions \cite{cordova-deluca-t-dso6}). In particular they are prominent ingredients in the AdS$_4\times \mathrm{CY}_6$ proposal  \cite{dewolfe-giryavets-kachru-taylor}.

As we mentioned in the introduction, it was found in \cite[Sec.~7]{saracco-t} that the O6 singularity could become smooth upon turning on $F_0\neq 0$. The curvature and dilaton no longer diverged; the singularity near the origin was replaced with a smooth $S^2$, on which the involution defining the O6 acted without a fixed locus. As we mentioned in the introduction, this is not to be thought of as a quantum resolution of the O6 singularity, since we are deforming it by the additional field $F_0$; the mechanism is very similar to one in field theory, where a Chern--Simons coupling can smooth out a moduli space singularity \cite{saracco-t-torroba}.\footnote{It is also important to stress that the local behavior found in \cite[Sec.~7]{saracco-t} cannot be extended to an asymptotically flat solution, since flat space is not a solution when $F_0\neq 0$.}

This phenomenon was found to happen under two conditions: assuming a certain symmetry that one would expect near the O6-plane; and working with a ``generic'' branch of the supersymmetry equations, where $F_6\neq 0$. However, as mentioned in the introduction, this branch has an obstruction that prevents it from being compact \cite{junghans-dgkt,marchesano-palti-quirant-t}. So we will analyze the ``special'' $F_6=0$ branch here with similar techniques as in \cite[Sec.~7]{saracco-t}, to see if we find again that the O6 singularity gets smoothed out.\footnote{This was claimed to have no solution in \cite[Sec.~7.5]{saracco-t}, but that subsection contained a mistake, as already pointed out in \cite{marchesano-palti-quirant-t}.}

As we mentioned, the fate of O6 singularities is very important in the AdS$_4\times \mathrm{CY}_6$ solutions. These were originally found by an effective $d=4$ supergravity approach, complemented by some $d=10$ intuition \cite{dewolfe-giryavets-kachru-taylor}. No reason was found for the EFT to break down here, but skepticism (see for example \cite{mcorist-sethi}) focused on whether an O6 might somehow be inconsistent with $F_0$, and on the fact that a $d=10$ uplift appeared to require smearing the O6 \cite{acharya-benini-valandro}. The first concern was partially answered by \cite[Sec.~7]{saracco-t}; since then, O6 singularities with $F_0\neq 0$ have appeared in other contexts, such as AdS$_7$ solutions and their descendants \cite{afrt,cremonesi-t,apruzzi-fazzi,rota-t}, undergoing some holographic checks \cite{apruzzi-fazzi}.\footnote{A related concern is that $F_0$ cannot be introduced in the world-sheet in the NSR formalism; however this is resolved in the Berkovits formalism  \cite{berkovits} (see for example \cite{benichou-policastro-troost} for a concrete application).} 

The second concern about the smearing was partially addressed more recently. These solutions are supposed to exist in the limit
\begin{equation}\label{eq:lim}
	g_s \sim R^{-3} \sim \mu \equiv \sqrt{-\Lambda/3} \sim N^{-3/4}\,.
\end{equation}
Here $N\in \mathbb{Z}$ is related to the $F_4$ flux quanta, and $g_s$ is a typical value for $\ee^\phi$ over the internal space. Using $g_s$ as an expansion parameter, the equations of motion and supersymmetry simplify considerably; using this, approximate solutions were found at the first nontrivial order in \cite{junghans-dgkt,marchesano-palti-quirant-t}. These have vanishing internal six-form flux, $F_6=0$, as we anticipated.

The approximation works well in most of the internal space, but it breaks down at a distance $\lesssim g_s l_s$ from the O6-planes: in that region, the higher orders appear to get larger than the first. This is comparable with the distance where the supergravity approximation itself breaks down; so in this sense the limit (\ref{eq:lim}) would seem to achieve all that could be hoped to obtain from supergravity. However, if a local analysis in this region showed a smooth-O6 deformation as in \cite[Sec.~7]{saracco-t}, we would end up with a solution we can trust everywhere, boosting our confidence in the validity of the AdS$_4\times \mathrm{CY}_6$ solutions. 

Having hopefully convinced the reader that the fate of the O6 singularity with $F_0 \neq 0 $ is important for scale separation, we will devote the rest of this section to its study.


\subsection{Supersymmetry} 
\label{sub:susy}

In this section we will work in string-frame variables. 
The supersymmetry equations for type II supergravity can be analyzed conveniently in terms of pure forms $\Phi_\pm$, associated bilinearly to the internal spinorial parameters $\eta^a_\pm$. In IIA they should satisfy the equations  \cite{gmpt2,gmpt3,t-reform}\footnote{A variety of normalizations is used in the literature for these forms; here we use $(\bar \Phi_\pm, \Phi_\pm)= \ee^{6A-2 \phi}$ as in \cite{saracco-t,marchesano-palti-quirant-t}, where $(\alpha, \beta)\mathrm{vol}_6\equiv (\alpha \wedge \lambda(\beta))_6$, ${}_6$ denoting taking the six-form part and $\lambda(\beta_k)\equiv (-1)^{\lfloor k/2 \rfloor} \beta_k$. Notice also that our $\dd_H$-closed RR forms $F$ were denoted by $G$ in \cite{marchesano-palti-quirant-t}.}
\begin{equation}\label{eq:psp}
	\dd_H \Phi_+ = -2\mu \ee^{-A} \mathrm{Re} \Phi_- \, ,\qquad {\mathcal J}_+ \cdot \dd_H(\ee^{-3A} \mathrm{Im} \Phi_-) = -5 \mu \ee^{-4A} \mathrm{Re} \Phi_+ + F\,.
\end{equation}
Here $\mu \equiv \sqrt{-\Lambda/3}$, $\dd_H\equiv \dd - H \wedge$,  $F=\sum F_k$ is the sum of all internal RR field strengths, and ${\mathcal J}_+ \cdot$ is an algebraic operator associated to $\Phi_+$ (see \cite{t-reform,saracco-t} for more details).

The forms $\Phi_\pm$ have a different expression depending on the angle $\psi$ between the spinors, defined by $|\eta^{2\,\dagger}_+ \eta^1_+|= || \eta^1_+|| || \eta^1_+|| \cos \psi$. 
\begin{itemize}
	\item When $\psi=0$ everywhere, the $\eta^a$ are parallel; when $F_0\neq 0$, $\Lambda<0$ this leads to constant $A$ and $\phi$ \cite{lust-tsimpis}, and so cannot be relevant for configurations with internal sources. 
	\item When $\psi=\pi/2$ everywhere, the $\eta^a$ are orthogonal; this is inconsistent with $\Lambda<0$ altogether \cite{bovy-lust-tsimpis,caviezel-kors-lust-tsimpis-zagermann}. 
	\item So we need to consider case where $\psi$ is generic. Here the forms $\Phi_\pm$ can be expressed in terms of a complex one-form $v$, a real two-form $j$, a complex two-form $\omega$, satisfying the $\mathrm{SU}(2)$-structure relations
	\begin{equation}\label{eq:su2}
		\iota_v j = \iota_v \omega = \iota_v \bar \omega=0 \, ,\qquad j \wedge \omega = \omega \wedge \omega= 0 \, ,\qquad \omega \wedge \bar \omega = 2 j^2\,.
	\end{equation}
	It is also useful to define
	\begin{equation}\label{eq:Jpsi}
		J_\psi \equiv \frac1{\cos\psi}j + \frac \ii{2 \tan^2 \psi } v\wedge \bar v \,,\qquad 
		\omega_\psi \equiv \frac1{\sin \psi} \left({\rm Re} \omega + \frac \ii{\cos\psi} {\rm Im} \omega\right)\,,
	\end{equation}    
	in terms of which $\Phi_+ = \ee^{3A-\phi} \cos\psi \exp[-i J_\psi]$, and $\Phi_- =\ee^{3A-\phi} \cos\psi\, v \wedge \exp[i \omega_\psi]$.
\end{itemize}

The consequences of (\ref{eq:psp}) were spelled out for this latter case in \cite[Sec.~5]{saracco-t}. The results depended crucially on a parameter $\theta$, defined as the phase of the zero-form part of $\Phi_+$:
\begin{equation}
	(\Phi_+)_0 =\ee^{3A-\phi + \ii \theta}\cos \psi\,.
\end{equation}
The two sub-cases $\theta\neq 0$ and $\theta=0$ have to be analyzed separately; they lead respectively to $F_6\neq 0$ and $F_6=0$. 

It was argued in \cite{junghans-dgkt,marchesano-palti-quirant-t} that $\theta=0$ (leading to $F_6=0$) is the correct case to consider. Indeed we now give an argument that $F_6=0$ should be the case for any compactification with an internal three-form source $\delta_3$ that is Poincar\'e dual to a cycle non-trivial in homology. The equations of motion for the internal fluxes $F_4$, $F_6$ are
\begin{equation}
	\dd (\ee^{4A} * F_4) + H \ee^{4A} * F_6 = 0\, ,\qquad
	\dd (\ee^{4A} * F_6)=0 \,.
\end{equation}
The second implies $\ee^{4A} * F_6$ is a constant. Suppose it is non-zero; then the first implies $H$ is exact. But then the Bianchi identity
\begin{equation}
	\dd F_2 - H  F_0 = \delta_3
\end{equation}
says $\delta_3$ is exact, in contradiction with the assumption.

So we restrict from now on to the $\theta=0$, $F_6=0$ branch. Here (\ref{eq:psp}) reduce to \cite[Sec.~5.2]{saracco-t}
\begin{subequations}\label{eq:revdJpsi}	
\begin{align}
	\label{eq:revt0}    
	&{\rm Re} v= - \frac{\ee^A}{2 \mu}\left(3 \dd A - \dd \phi - \tan \psi \dd \psi\right) = -\frac{\ee^A}{2\mu} \dd\log(\cos\psi \ee^{3A-\phi}) \, ,\\
    \label{eq:dJpsit0}
    &\dd(\ee^{3A-\phi} \cos \psi J_\psi)=0 \, 
\end{align}
\end{subequations}
and
\begin{subequations}\label{eq:flux-t0}
\begin{align}
	\label{eq:Ht0}
	&H= \hat H + \dd(\tan\psi \mathrm{Im} \omega) \, ,\qquad 
	\hat H= 2\mu \ee^{-A} {\rm Re} (\ii v \wedge \omega_\psi) \, ,\qquad
	F= \ee^{\tan \psi \mathrm{Im} \omega \wedge} \hat F\,;\\
	\label{eq:F0t0}
	& F_0 = -J_\psi\cdot
    \dd (\cos \psi \ee^{-\phi}{\rm Im} v)
    + 5 \mu \cos \psi \ee^{-A-\phi} \,,\\  
	\label{eq:F2t0}
	& \hat F_2 = -J_\psi\cdot \dd\, {\rm Im} (\ii \cos \psi \ee^{-\phi} v \wedge \omega_\psi) - 2 \mu \frac{\sin^2 \psi}{\cos \psi} \ee^{-A-\phi} {\rm Im} \omega_\psi\,,\\
	\label{eq:F4t0}
& \hat F_4 = J_\psi^2\left[ \frac12 F_0 - \mu \cos \psi \ee^{-A-\phi}\right] + J_\psi \wedge \dd\, {\rm Im} (\cos \psi \ee^{-\phi} v) \,,\\
& \hat F_6 = 0 \,.
\end{align}
\end{subequations}

To make sure one has a supersymmetric solution, one needs to solve (\ref{eq:revdJpsi}), and to make sure that the Bianchi identities are satisfied, which away from sources reduce to 
\begin{equation}
	\dd \hat H = 0 \, ,\qquad \dd_{\hat H} \hat F=0\,.
\end{equation}

One should finally make sure all fields and pure forms transform correctly under any O$p$. This is usually defined as
\begin{equation}\label{eq:OmegaRp}
	\begin{array}{cc}
		\Omega R_p \quad &\text{if }p=0,\,4,\,8,\,;\\
		\Omega (-1)^{F_\mathrm{L}} R_p \quad &\text{if }  p=2,\,3,\,6,\,7\,. 
	\end{array}
\end{equation}
where $R_p$ is an involution reversing $9-p$ coordinates, and leaving $p+1$ invariant (including time); of course in IIA only even $p$ are relevant. One should then impose \begin{equation}\label{eq:O}
\begin{split}
	&R_p^* \phi = \phi \, ,\qquad R_p^* g = g \, ,\qquad R_p^* B= - B \, ,\qquad R_p^* F_k = -(-1)^{\lfloor p/2 \rfloor + \lfloor k/2 \rfloor} F_k\,,\\
	&R^*_p \Phi_\pm= -(-1)^{\lfloor\frac p2 \rfloor}\lambda(\Phi_\pm)\,,
    \qquad
    R^*_p \Phi_\mp =
    (-1)^{\lfloor\frac{p-1}2\rfloor}\lambda(\bar\Phi_\mp)\quad \overset{\text{\tiny{IIA}}}{\text{\tiny{IIB}}}\,,
\end{split}
\end{equation}
where $R_p^*$ denotes pull-back; in IIA the upper sign is relevant.


\subsection{The emergent symmetry} 
\label{sub:sym}

We call $x^i$ the three internal coordinates parallel to the O6, and  $y^i$ the transverse coordinates. So the involution defining the O6 maps  $(x^i,y^i)\to (x^i,-y^i)$.

Near the O6, we expect the solution to become invariant under translations of the $x^i$. We might also expect invariance under rotations of the $x^i$ and of the $y^i$. However, following \cite{saracco-t} we will relax this and impose symmetry only under simultaneous rotations of the $x^i$ and $y^i$. The total symmetry group is then
\begin{equation}\label{eq:iso}
	\mathrm{ISO}(3)\,.
\end{equation}
This still implies that all the functions $A$, $\phi$, $\psi$ in the equations will need to depend only on the transverse radial coordinate $r\equiv \sqrt{y^i y^i}$, as is reasonable to expect. On the one-forms, it allows 
\begin{equation}\label{eq:1f}
	\omega_{1,0}\equiv y^i \dd y^i \, ,\qquad \omega_{1,1} \equiv y^i \dd x^i\,.
\end{equation}
Separate rotational invariance would allow only $\omega_{1,0}$.

A basis of two-forms invariant under the symmetry (\ref{eq:iso}) is 
\begin{equation}\label{eq:2f}
	\begin{split}
	\omega_{2,0}&\equiv\epsilon_{ijk} y^i \dd y^j \wedge \dd x^k \, ,\qquad
		\omega_{2,1}\equiv \epsilon_{ijk} y^i \dd y^j \wedge \dd y^k \, ,\qquad
		\omega_{2,2}\equiv \epsilon_{ijk}y^i \dd x^j \wedge \dd x^k\,
		\\
		&\omega_{2,3}\equiv \omega_{1,0}\wedge \omega_{1,1} \, ,\qquad
		\omega_{2,4}\equiv \dd x^i \wedge \dd y^i\,.
	\end{split}
\end{equation}
There are also eight invariant three-forms $\omega_{3,i}$ \cite[(A.3)]{saracco-t}.

As in \cite{saracco-t}, we will postulate this symmetry and see where it leads. So for example the one-form $v$ will then be a linear combination of (\ref{eq:1f}) alone, and not of more general forms such as $\dd x^1$, which would be invariant under both $y^i$ rotation and $x^i$ translation but not under simultaneous rotation. Likewise, $j$, $\omega$ will be a linear combination of (\ref{eq:2f}). 

We also need to impose (\ref{eq:O}) for $p=6$, which imply
\begin{equation}
	R_6^* v= \bar v \, ,\qquad R_6^* j= - j \, ,\qquad R_6^* \omega = \bar \omega\,,
\end{equation}
where $R_6: y^i\to -y^i$. All in all this gives \cite[(7.18)]{saracco-t}
\begin{equation}\label{eq:vjo-coeff}
	v= v_\mathrm{r} \omega_{1,0} + \ii v_\mathrm{i} \omega_{1,1}\, ,\qquad
	j= \sum_{i=1}^4 j_i \omega_{2,i} \, ,\qquad \omega= a_0 \omega_{2,0}+\sum_{i=1}^4 \ii a_i \omega_{2,i}\,,
\end{equation}
where all the coefficients are real.

The algebraic conditions (\ref{eq:su2}) become the quadratic equations
\begin{equation}\label{eq:alg}
\begin{split}
	&j_4= j_3 r^2 \,,\qquad 
	a_4= a_3 r^2 \,,\qquad
	a_2 j_1 + a_1 j_2 = \frac12 a_3 j_3 r^2\,, \\
	&a_3^2 r^2 - 4 a_1 a_2 = a_0^2=
		j_3^2 r^2 - 4 j_1 j_2 \,.
\end{split}
\end{equation}
Strictly speaking, these follow already from the conditions in (\ref{eq:su2}) involving $j$ and $\omega$ alone. The conditions in (\ref{eq:su2}) involving the contraction $\iota_v$ are to be read as constraints on the metric, since they involve $g^{mn}v_n$; in general they are automatically satisfied by taking the vielbein to be of the form
\begin{equation}\label{eq:vielbein}
	\{ \mathrm{Re} v,\, \mathrm{Im} v,\, e^a\}
\end{equation}
where $e^a$, $a=1,\,\ldots,\,4$ are such that $j= e^1 \wedge e^3 + e^2 \wedge e^4$, $\omega= (e^1+ \ii e^3) \wedge (e^2 + \ii e^4)$. However, in our present setup, the invariant vector fields are linear combinations of $y^i \partial_{y^i}$ and $y^i \partial_{x^i}$, and the contractions of these with (\ref{eq:2f}) had better be zero. Indeed these are also satisfied thanks to the first two in (\ref{eq:alg}), so all is well.

The metric is determined by the vielbein (\ref{eq:vielbein}). The $e^a$ are not always easy to find, but alternatively one may proceed as follows. The sum $g_4\equiv \sum_{a=1}^4 e^a e^a$ is a rank-four symmetric tensor. At each point, it defines a positive-definite metric on the four-dimensional quotient space
\begin{equation}\label{eq:TM/v}
	TM_6/\mathrm{Span}(v,\bar v)
\end{equation}
determined by $j$ and $\omega$ as for an $\mathrm{SU}(2)$-structure in four dimensions. Inspired by this, we introduce vector fields $E^a$, $a=1,\,\ldots,\,4$ such that $\iota_{E^a}e^b= \delta_a^b$, $\iota_{E^a} v= \iota_{E^a}\bar v=0$. The matrix $\Pi_4 \equiv \sum_{i=1}^4 e^a \otimes E_a$ is a projector, $\Pi_4^2 = \Pi_4$; using our symmetry assumptions, it is found to be $\Pi_4 = 1_6 - y^i y^j (\dd y^i \otimes \partial_{y^j} + \dd x^i \otimes \partial_{x^j})$. We also introduce the bi-vector $(\mathrm{Re} 	\omega)^{-1}\equiv - E^1 \wedge E^2 + E^3 \wedge E^4$, and 
\begin{equation}
	I_4 = -\mathrm{Im} \omega (\mathrm{Re} 	\omega)^{-1}
\end{equation}
which satisfies $I_4^2= - \Pi_4$ and is thus an almost complex structure on the spaces (\ref{eq:TM/v}). In our case, $\mathrm{Re} \omega= a_0 \omega_0$, and $(\mathrm{Re} \omega)^{-1}$ is easy to guess: it should be invariant under (\ref{eq:iso}) and even under $R_6$, which fixes it to be $-(r^2 a_3)^{-1}\epsilon^{ijk}y^i \partial_{x^j} \wedge \partial_{y^k}$. Finally then we compute $(g_4)_{mn} = - I_m{}^p j_{pn}$. 

After some simplification, the final result for the internal metric can be written as an $S^2$-fibration over $\mathbb{R}^3$. Defining $\hat{y}^i\equiv y^i/r$:
\begin{align}
	\nonumber
	\dd s^2_6 &=\frac{r^2 v_\mathrm{r}^2}{\tan^2 \psi}\dd r^2 + 2 a_0^{-1}r^4 (a_1 j_3 - a_3 j_1) D \hat{y}^i D \hat{y}^i\\ 
	\label{eq:met0}
	&+\left(\frac{a_0^3}{a_3 j_2 - a_2 j_3}(\delta_{ij}-\hat{y}^i \hat{y}^j) + \frac{r^2 v_\mathrm{i}^2}{\tan^2 \psi} \hat{y}^i \hat{y}^j\right) \dd x^i \dd x^j\,;\\
	\nonumber
	&D \hat{y}^i \equiv \dd \hat{y}^i+ r^{-2} \frac{a_2j_1 - a_1 j_2}{a_1 j_3 - a_3 j_1} \epsilon^{ijk} \hat{y}^j \dd x^k\,.
\end{align}
(The term $D \hat{y}^i D\hat{y}^i$ is a fibred version of $\dd s^2_{S^2}= \dd y^i \dd y^i$.) The fibration is topologically trivial, so it can also be written as an $\mathbb{R}^3$ fibred over $S^2$ as in \cite[Sec.~7.3.2]{saracco-t}:
\begin{align}
	\nonumber
	\dd s^2_6 &=\frac{r^2 v_\mathrm{r}^2}{\tan^2 \psi} \dd r^2 +\frac{r^2}2 \frac{a_0^3}{a_1 j_3 - a_3 j_1}\dd s^2_{S^2}\\
	\label{eq:met}
	&+r^2\left(2 a_0^{-1}(a_3 j_2-a_2 j_3) (\delta_{ij}- \hat{y}^i \hat{y}^j) + \frac{v_\mathrm{i}^2}{\tan^2 \psi} \hat{y}^i \hat{y}^j \right) D x^i D x^j\,;\\
	\nonumber& D x^i \equiv \dd x^i + r^2 \frac{a_2 j_1 - a_1 j_2}{a_2 j_3 - a_3 j_2}\epsilon^{ijk}\hat{y}^j \dd \hat{y}^k \,.
\end{align}
The advantage of (\ref{eq:met0}) is that the fiber metric is independent of the base coordinates; however, in what follows (\ref{eq:met}) will result in simpler expressions. From both perspectives, the invariance under simultaneous rotations (the $\mathrm{SO}(3)$ subgroup of (\ref{eq:iso})) can be seen as the lift of a base isometry to that of the total space, as usual for a fibration. 

Positive-definiteness of (\ref{eq:met0}) requires $\beta_{13} \equiv a_0(a_1 j_3 - a_3 j_1)$, while for (\ref{eq:met}) it requires $\beta_{32} \equiv a_0(a_3 j_2-a_2 j_3)$. The two conditions are equivalent because (\ref{eq:alg}) implies $r^2\beta_{13} \beta_{32}= \frac14 a_0^2 + a_0^{-2} (a_2 j_1 - a_1 j_2)^2>0$.

The setup in this section is very similar to that in \cite{rota-t}, where the $S^2$ was fibered on a three-dimensional (locally) maximally symmetric space $\Sigma_3$. A symmetry similar to (\ref{eq:iso}) also appears there; the basis of forms \cite[(3.16)]{rota-t} is linearly related to our (\ref{eq:2f}). In particular the case $\Sigma_3= \mathbb{R}^3$ is directly related to the present section, but more generally one expects the curvature of $\Sigma_3$ to become irrelevant in the local limit we are interested in here. However, some of the $j_i$, $a_i$ coefficients in (\ref{eq:alg}) were set to zero in \cite{rota-t} because of the particular applications that motivated that paper.


\subsection{The differential system} 
\label{sub:sys}

The differential equations in Sec.~\ref{sub:susy} can be reduced using the symmetry in Sec.~\ref{sub:sym}. This is the same method used in \cite[Sec.~7.3]{saracco-t}, but as we will see the results are quite different.

We start our analysis from the three-form equation (\ref{eq:dJpsit0}). 
Two of its components can be readily solved, giving
\begin{equation}\label{eq:threeForm1}
	j_2=0\,,\quad j_1 = c_1  \frac{\ee^{\phi - 3 A}}{r^3}\,,
\end{equation}
where $c_1$ is an unfixed constant, which will interpret geometrically momentarily.
After imposing \eqref{eq:threeForm1},  only a single non-trivial equation is left:
\begin{equation}
	j_4  (3 A' - \phi') + j_3 r + j_4' + rv_0 v_\mathrm{i} \frac{\cos^3\psi}  {\sin^2  \psi}  = 0\;.
\end{equation} 
From (\ref{eq:alg}) it now also follows that $a_0=r j_3$.
We then have to impose the Bianchi equations for the fluxes. 
From $\dd \hat{H} = 0$ we obtain a single independent equation:
\begin{equation}
	a_0  v_\mathrm{i}  (rA' + r \psi' \cot  \psi  - 3) - r (a_0 v_\mathrm{i})'  + 2 a_2 v_\mathrm{r}
\sec  \psi  = 0\;.
\end{equation}
The Bianchi identities for $\hat{F}$ are more involved. Expanding \eqref{eq:F0t0}, we obtain that $F_0$ is given by
\begin{equation}
	F_0  =  5 \mu \ee^{- A - \phi} \cos  \psi  + \frac{2 v_\mathrm{i} \ee^{- \phi} \cos^2
	 \psi }{j_3 r^2} - \frac{\ee^{- \phi} \sin^2  \psi}{r v_\mathrm{r}} \left(-\frac1r + \phi' +\tan \psi \, \psi' -\frac{v_\mathrm{i}'}{v_\mathrm{i}}\right)\,.
\end{equation}
This is a first order differential equation that has to be solved.
To organize the other Bianchi identities we first define
\begin{equation}\label{eq:deff}
  \hat{F}_2 = \sum_{i = 0}^4 f_{2 i} \omega_{2 i}, \qquad \hat{F}_4 = \sum_{i
  = 0}^4 f_{4 i} \omega_{4 i}\;.
\end{equation}
Upon expading \eqref{eq:F2t0} and \eqref{eq:F4t0} we read $f_{4 0} = f_{20} = 0 $.
In terms of the newly introduced variables, $\dd_{\hat{H}} \hat{F}_2 = 0$ gives 4 non-trivial differential equations:
	\begin{subequations}
		\begin{align}
			0 & =  2 r (4 a_1 \ee^{- A} F_0 \mu rv_\mathrm{r} \csc (2 \psi) + f_{21}') + 6 f_{21}
			\label{eq:dF2-1}\;,\\
			0 & =  a_0 \ee^{- A} F_0 \mu r^2 v_\mathrm{i} \csc  \psi  + f_{22} \label{eq:dF2-2}\;,\\
			0 & =  4 a_3 \ee^{- A} F_0 \mu r^2 v_\mathrm{r} \csc (2 \psi) + \frac{f_{24}'}{r} +
			f_{23} \label{eq:dF2-3}\;,\\
			0 & =  4 a_2 F_0 \mu rv_\mathrm{r} \csc (2 \psi) - a_0 F_0 \mu rv_\mathrm{i} \csc  \psi  +
			\ee^A f_{22}' \;;\label{eq:dF2-4}
		\end{align}
	\end{subequations}
while $\dd_{\hat{H}} \hat{F}_4 = 0$ produces a single differential equation: 
\begin{equation}\label{eq:dHF4}
  0  =  4 \ee^{- A} \mu r^2 v_\mathrm{r}  (2 a_2 f_{21} + 2 a_1 f_{22} - a_3 f_{24})
  \csc (2 \psi) + 2 rf_{41}' + \frac{f_{44}'}{r} + 8 f_{41} \;.
\end{equation}

Generically, any equation where an $f_{i j}'$ appears is in danger of being a second order differential equation
once the definition \eqref{eq:deff} of $f_{i j}$ is substituted.
However, an explicit computation shows that this is not the case for the particular combination in \eqref{eq:dHF4}.
Similarly, \eqref{eq:dF2-4} can be reduced to a first order equation by substituting $f_{22}$ from \eqref{eq:dF2-2}.

As a result, only \eqref{eq:dF2-1} and \eqref{eq:dF2-3} would become second order
equations if we subsitute $f_{21}$ and $f_{24}$ right away. To avoid this, we keep these two components of the fluxes as variables.
Finally, \eqref{eq:revt0} gives an extra first order equation.

Summing up, at this stage we have 11 first order differential equations and two remaining algebraic equations,
\begin{equation}\label{eq:algRem}
	0 =r^2 \left(a_3^2-j_3^2\right)-4 a_1 a_2 \, ,\qquad 0 = a_3 j_3 r^2-\frac{2 a_2 c_1 \ee^{\phi -3 A}}{r^3}\,,
\end{equation}
for the 11 unknown functions
\begin{equation}
a_1,\; a_2,\;a_3\;,j_3\;, v_\mathrm{r},\; v_\mathrm{i},\;f_{21},\;f_{24},\;A,\;\psi,\;\phi \:.
\end{equation}

The system can thus be further reduced. 
Without loss of generality, we can assume $a_2 \neq 0$ and $j_3 \neq 0$ and solve the remaining algebraic equations \eqref{eq:algRem} by setting
\begin{equation}
	a_3 = \frac{2 c_1 \ee^{-3A+\phi} a_2}{r^5 j_3},\qquad a_1 = \frac{r^2 \left(a_3^2-j_3^2\right)}{4 a_2}\;.
\end{equation}
Plugging this back in the differential system we find an extra algebraic combination, which can be used to solve for $f_{24}$ as 
\begin{equation}
	f_{24} = 	-2 c_1\frac{\ee^{-4 A} v_\mathrm{i}}{\sin \psi} \left(F_0 \mu  \ee^{\phi } + \frac{a_2 \ee^A \cos  \psi }{j_3^2 r^5} \right)\;.
\end{equation}
We then find that three more of the differential equations are redundant, leaving us with seven equations for eight variables.
This apparent tension is resolved once we account for possible redefinitions of the coordinate $r$.
This can be more easily seen introducing the variables
\begin{equation}
	\rho_0 \equiv v_\mathrm{r} r,\qquad \rho_1 \equiv r^3 j_3 \ee^{3A-\phi},\qquad \rho_2 \equiv r a_2,\qquad \rho_3 \equiv r v_\mathrm{i},\qquad\tilde{f}_{21}\equiv r^3 f_{21}\;,
\end{equation}
where the last four quantities now transform as functions under a change of the coordinate $r$. 
As a result of these redefinitions, the remaining system becomes autonomous (i.e.~without an explicit $r$ dependence) and it is left invariant by the transformation
\begin{equation}\label{eq:gaugeTrans}
	\dd r \to \ee^{Q}\dd r\,,\qquad \rho_0 \to \ee^{-Q}\rho_0\;,
\end{equation}
where $Q$ is an arbitrary function of $r$.
Another way to understand this property is by looking at the six-dimensional metric;
after the chain of substitutions in this section, (\ref{eq:met}) becomes
\begin{equation}\label{eq:metricRed}
	\dd s^2_6 = \left( - 2 \rho_2 (\delta_{i j}- \hat{y}^i \hat{y}^j)  + \frac{\rho_3^2}{\tan^2  \psi }  \hat{y}^i \hat{y}^j  \right) D x^i D x^j
	 +\frac{\rho_0^2}{\tan^2  \psi } \dd  r^2 -\frac{\rho _1^2 \ee^{2 \phi -6 A}}{2 \rho _2} \dd  s^2_{S^2}\,,
\end{equation}
with $Dx^i\equiv \dd  x^i + \frac{c_1}{\rho_1} \varepsilon^{i}_
{j k} \hat{y}^j \dd  \hat{y}^k$. We can immediately see that \eqref{eq:gaugeTrans} cancels from the metric.
Moreover, the constant $c_1$ controls the internal fibration, turning it off completely when it vanishes.

To summarize so far, the freedom in \eqref{eq:gaugeTrans} to fix $\rho_0$ gives the same number of equations and variables.
In our solutions below, we will fix the gauge by choosing $\rho_0$ such that the $g_{rr}$ component of the metric \eqref{eq:metricRed} reduces to $\ee^{-2A}$:
\begin{equation}\label{eq:gauge}
	g_{rr}= \frac{\rho_0^2}{\tan^2 \psi}= \ee^{-2A}\,.
\end{equation}
This is the same gauge in which the massless solution is usually written, allowing for a direct comparison of the massless limit.

All in all, we are left with a system of seven  first order differential equations for seven variables:
\begin{subequations}\label{eq:systemFinal}
	\begin{align}
		\rho_1'&=-\ee^{2 A-\phi } \rho_3 \frac{\cos^2\psi}{\sin \psi}\,,\label{eq:systemFinal-a}\\
		\frac{\rho _2'}{\rho_2}&= \phi' + \cot  \psi\, \psi' -\frac{\rho _3'}{\rho _3}
		+2 \mu\ee^{-2 A}\tan  \psi   - F_0 \mu  \ee^{-5 A + \phi}\frac{\rho _1 \rho _3}{\rho_2}  \cot  \psi \,, \label{eq:systemFinal-b}\\
		\frac{\rho _3'}{\rho_3}&= 4 A'-\phi'+ \cot  \psi\, \psi ' + \ee^{2 A-\phi }\frac{\rho_3^2 \cos^4 \psi +2 \rho_2 \sin^2 \psi}{\rho_1 \rho_3 \sin \psi \cos^2 \psi}\,,\label{eq:systemFinal-c}\\
		\tilde{f}_{21}'&=\frac{ F_0 \mu \ee^{-8 A}}{2 \rho _1^2 \rho _2  \cos^2 \psi  }\left(\rho _1^4 \ee^{2 \phi }- c_1^2 \ee^{6 A}\rho _2^2\right) \,,\label{eq:systemFinal-d}\\
		\psi'&=\cot  \psi  \left(3 A'-\phi '\right)+2 \mu \ee^{-2 A} \,, \label{eq:systemFinal-e}\\
		\phi'&=\frac{\rho _3'}{\rho_3}-\tan  \psi\,\psi ' + \frac{\ee^{-2 A}}{\sin\psi} \left(F_0 \ee^{A+\phi }-5 \mu  \cos  \psi \right)-2 \ee^{2 A-\phi } \frac{\rho _3 \cos^2 \psi}{\rho_1 \sin \psi}  \,,\label{eq:systemFinal-f}\\
		4 A'&= \frac{2F_0 \mu}{\sin \psi} \rho_3(\ee^{-5A + 2 \phi} \rho_1 \rho_2^{-1} + \ee^A c_1^2 \rho_1^{-3}\rho_2) -2\ee^{-2A}\frac{\sin \psi}{\cos^2 \psi}(F_0 \ee^{A + \phi}- \mu \cos \psi)  \label{eq:systemFinal-g}\\
		\nonumber & +2 \ee^{2A- \phi}\rho_1^{-1}\rho_3 \sin \psi + 2c_1^2 \ee^{8A-3 \phi}\frac{\rho_2^2 \rho_3}{\rho_1^5 \sin \psi} +4 \ee^{5A-\phi}\frac{\tilde f_{21}\rho_2}{\rho_1^2 \cos \psi}\,. 
\end{align}
\end{subequations}
This system is left invariant by the independent rescalings
\begin{equation}
	r \rightarrow e^{k_0} r, \qquad A \rightarrow A + k_1, \qquad \phi
\rightarrow \phi + k_2, \quad \rho_3 \rightarrow e^{k_3} \rho_3\;,
\end{equation}
where $k_i$ are real constants; the other functions and the parameters transform as
\begin{equation}
	\begin{split}
		&\rho_1 \rightarrow \ee^{k_0 + 2 k_1 - k_2 + k_3}\rho_1\,, \qquad \rho_2 \rightarrow
	\ee^{2 k_3}\rho_2\,, \qquad \tilde{f}_{21} \rightarrow \ee^{k_0 - k_1 - k_2}
	\tilde{f}_{21}\,,\\
	&F_0 \rightarrow \ee^{- k_0 - k_1 - k_2} F_0\,, \qquad \mu \rightarrow \ee^{- k_0 +
2 k_1} \mu\,, \qquad c_1 \rightarrow \ee^{2 k_0 + k_1 - k_2} c_1 \,.
	\end{split}
\end{equation}
These can be used to achieve small curvature and dilaton, and to adjust the flux quanta so that they are integer. However, they cannot be used to achieve parametric separation of scales: interestingly, the combination $\text{diam}^2_{\bar{g}} |\Lambda|$,
which appears in the bounds for $\frac{m^2_k}{|\Lambda|}$ such as \eqref{eq:lowerBound} and Theorem \ref{thm:boundDiam}, does not rescale.


\subsection{Boundary conditions} 
\label{sub:bc}

As usual, the next step is to study the possible boundary conditions for this system. For example, the internal space can end where the $S^2$ shrinks; the latter can combine with the $\dd r^2$ term in the metric to give a smooth space. More generally, one may look for singularities which have a known interpretation as the back-reaction of O-planes and D-branes. To find these local behaviors, we used two techniques. 

The first is a non-linear analogue of the Frobenius method: it consists in postulating a power series expansion, often with fractional or negative exponents, suggested by a physical behavior one wants to achieve. The coefficients are then determined order by order by the differential equations. 

The second is backward numerical evolution: one takes a random value for all variables, and evolves the system (\ref{eq:systemFinal}) numerically backwards in $r$. The evolution stops at a value $r=r_0$ where some of the functions diverge or go to zero. The leading powers of the variables near $r_0$ can be determined by inspecting the numerical data.

These two techniques are complementary: some boundary conditions of obvious physical significance can be obtained easily by power series expansion, but are not easily found numerically. This indicates that they are not attractors for the system, but require fine tuning. On the other hand, some numerical attractors can only be interpreted physically with some effort, or not at all. We summarized some notable local behaviors found with both methods in Table \ref{tab:bc};\footnote{In our gauge (\ref{eq:gauge}), the subleading powers are all separated by integers from the leading ones. This is unlike in \cite{rota-t}, which as we commented earlier studied a related system, and where many of the same singularities also appeared.}
let us go through some of them in more detail.

\begin{table}
\begin{center}
\begin{tabular}{cccccccc}
$\rho_1$ & $\rho_2$  & $\rho_3$  & $\tilde{f}_{21}$  & $\tan\psi$  & $\ee^\phi$  & $\ee^A$  & interpretation \\\hline\hline
$1$   & $0$  & $1$  & $3$  & $1$  & $0$  & $0$  & regular \\\hline
$0$   &$1/2$ &$-3/4$& $0$  &$-1/2$&$-1/4$&$-1/4$& partially smeared O4  \\\hline
$0$   &$-1/2$&$1/4$ & $0$  &$1/2$ &$-3/4$&$-1/4$& O6 \\\hline
$1$   &$1/2$ &$3/4$ & $0$  &$1/2$ &$3/4$ &$1/4$ & D6 \\\hline
$0$   &$0$   &$-1$  &$-1$  &$-1$  &$0$   &$0$   & boundary \\\hline
$0$   &$1/2$ &$0$   & $0$  &$-1/2$&$1/2$   &$0$ & $\sigma$ \\\hline
\end{tabular}
\end{center}
\caption{Some notable boundary conditions. The numbers denote the leading powers of the corresponding variables. The regular, D6-brane and O6-plane require fine-tuning, while the others occur as attractors.}
\label{tab:bc}
\end{table}

\subparagraph{Regular point.} As an example of the Frobenius method, the $S^2$ shrinks regularly if the functions behave as
\begin{align}\label{eq:reg}
	\nonumber&\rho_1 = \frac{b_2 b_3^3 F_0 r}{b_1(5 \mu + 3 b_1 b_3^2)}+O(r^3) \, ,\qquad
	\rho_2 = -\frac{b_2^2}{2b_1^2} + O(r^2) \, ,\qquad 
	\rho_3 = b_3 r + O(r^3)\, ,\qquad\\
	&\tilde{f}_{21}= \frac{\mu F_0 r^3}{3b_3^4} + O(r^5)\, ,\qquad
	\psi = b_1 r + O(r^3)\, ,\qquad\\
	\nonumber&\ee^\phi = \frac{5\mu +3 b_1 b_3^2}{b_3 F_0} + O(r^2)\, ,\qquad
	\ee^A = b_3 +O(r^2)\,,
\end{align}
where $b_i$ are three constants, and we took $c_1=0$. By (\ref{eq:metricRed}), the local metric is then, at leading order,
\begin{equation}
	\dd s^2_{10}\sim b_3^2 \dd s^2_{\mathrm{AdS}_4}+\frac{b_2^2}{2b_1^2}\dd x^i \dd x^i + \frac1{b_3^2}(\dd r^2 + r^2 \dd s^2_{S^2})\,.
\end{equation}
We indeed see that the parenthesis reconstructs an $\mathbb{R}^3$. The other fields are also smooth. This local behavior is not an attractor, so it is not found by numerical evolution without fine tuning. In any case, this behavior cannot look like an O6-plane from far away: the $S^2$ shrinks without a point charge. So this will not be relevant for the rest of the paper. 

\subparagraph{Boundary.} On the other hand, one may want to look for a different type of smooth behavior, more similar to that in \cite[Sec.~7]{saracco-t}. The idea is that the local metric would look like
\begin{equation}\label{eq:sar}
	\dd r^2 + \dd s^2_5(r)
\end{equation}
 around $r=r_0$, with the first derivatives  $\partial_r\dd s^2_5(r_0)=0$ and similar conditions for the other fields. This would represent an actual boundary for the space, but it could be glued to a second copy of the same solution to obtain a compact smooth space. (The numerical evolution might even continue automatically past such a value, in appropriate variables, but it is more common for it to stop as in \cite[Sec.~7]{saracco-t}.) We found numerical candidates for such points as numerical attractors: for an open set in the space of all initial values, backwards numerical evolution stops at such a point. This is denoted as ``boundary'' in Table \ref{tab:bc}.\footnote{$\tilde{f}_{21}$ diverges, but the corresponding component of the physical flux as defined in (\ref{eq:flux-t0}) is finite.} We were then able to find a corresponding local solution by the power series method. Unfortunately we found that the partial derivative $\partial_r \frac{\rho_3^2}{\tan^2 \psi}(r_0)\neq 0$: this is the coefficient of the term $(\hat{y}^i D x^i)^2$ in (\ref{eq:metricRed}). So this cannot be a smooth point. We then looked more broadly for behaviors of the type (\ref{eq:sar}), attractors or not; but in no case we were able to achieve full regularity.
 
A more general smooth boundary with $\psi\neq\frac{\pi}{2}$ can be directly excluded from the differential system \eqref{eq:systemFinal}.
Indeed, from the factor $-\frac{\rho_1^2 \ee^{2\phi-6A}}{2\rho_2}$ in front of the $\dd s^2_{S^2}$ term in the metric \eqref{eq:metricRed},
we see that a smooth boundary where the sphere does not shrink would require a non-vanishing $\rho_1$ with a zero derivative at $r = r_0$.
From \eqref{eq:systemFinal-a}, $\rho_1' = 0$ requires $\rho_3$ to vanish, since we are assuming $\psi\neq \frac{\pi}{2}$.
Thus, if $\rho_3$ were to vanish as a power series, the term $\frac{\rho_3'}{\rho_3}$ in \eqref{eq:systemFinal-f} would diverge as $\sim r^{-1}$ and it could not be compensated by any of the other terms in the equation, which by assumption would go to (possibly zero) constants.

It does happen sometimes that the numerical evolution stops at a point where $\psi$ and $\rho_3$ go to zero linearly and all other variables are finite, but again (\ref{eq:systemFinal-a}) shows that $\rho_1'\neq 0$.

These results already dash the hopes that motivated this section: had we found a regular boundary, we could have hoped to glue it to the solution perturbative in $g_s$ from \cite{junghans-dgkt,marchesano-palti-quirant-t}, completing it. In any case, we will carry on to describe other boundary conditions, to see what alternatives are available.

\subparagraph{O6-plane, D6-brane.} The local solution
\begin{align}\label{eq:o6}
	\nonumber&\rho_1 = b_1 +O(r) \, ,\qquad
	\rho_2 = -\frac{b_2}{\sqrt r} + O(r^{1/2}) \, ,\qquad 
	\rho_3 = b_3 r^{1/4} + O(r^{3/4})\, ,\qquad\\
	&\tilde{f}_{21}= \frac{-b_1^5 b_5^4 - \mu c_1^2 b_1^2 b_2 b_3 b_4^2 b_5^3 + 2 c_1^2 b_2^2 b_3 b_4^9 F_0^{-1}}{4 b_1^3 b_2 b_4^5 b_4^5} + O(r)\, ,\qquad\\
	\nonumber &\psi = \frac{b_5 F_0 \sqrt{r}}{b_4} + O(r^{3/2})\, ,\quad
	 \ee^\phi = b_5 r^{-3/4}+ O(r^{1/4})\, ,\quad
	\ee^A =b_4 r^{-1/4} +O(r^{3/4})\,
\end{align}
leads to the metric
\begin{equation}
	\dd s^2_6 \sim \frac1{\sqrt r}\left(b_4^2\dd s^2_{\mathrm{AdS}_4}+ \left(b_2 (\delta_{ij} - \hat{y}^i \hat{y}^j) + \frac{b_3^2 b_4^2}{b_5^2 F_0^2}\hat{y}^i \hat{y}^j\right)D x^i D x^j\right) + \sqrt{r} \left(\frac{\dd r^2}{b_4^2} + \frac{b_1^2 b_5^2}{b_2 b_4^6} \dd s^2_{S^2}\right)\,.
\end{equation}
This has the usual structure $h^{-1/2}\dd s^2_\parallel + h^{1/2} \dd s^2_\perp$ of D$p$-branes and O$p$-branes in the string frame, with $h\sim r$; there are three transverse directions, $p=6$.
The linear behavior can be interpreted as the local expansion of $h_\mathrm{O6}= 1- \frac{r_0}r$, $r_0=l_s g_s$ around the locus $\{r=r_0\}$, the boundary of the excluded hole around the source. 

The local behavior for a D6-brane can also be found with the power series method, but we will not give the details here; the leading powers can be found in Table \ref{tab:bc}.

Neither the O6 nor the D6 are attractors: they can only be found numerically by a great deal of fine tuning. This will play a role in the next section.

\subparagraph{Partially smeared O4.}
The following singularity is the one that occurs most often in numerical evolution: not only it is an attractor, its open basin of attraction appears to be the largest. We first found it numerically, and then reproduced it in a power series expansion as
\begin{align}\label{eq:o4}
	\nonumber&\rho_1 = b_1 +O(r) \, ,\qquad
	\rho_2 = - b_2 \sqrt r + O(r^{3/2}) \, ,\qquad 
	\rho_3 = b_3 r^{-3/4} + O(r^{1/4})\, ,\qquad\\
	&\tilde{f}_{21}=\tilde{f}_{21,0}(b_i) + O(r)\, ,\qquad\psi = \frac\pi2- b_6 \sqrt{r}+ O(r^{3/2})
	\, ,\\
	\nonumber &
	 \ee^\phi = b_5 r^{-1/4}+ O(r^{1/4})\, ,\quad
	\ee^A =b_4 r^{-1/4} +O(r^{3/4})\,\\
	\nonumber&\dd s^2_6 \sim \frac1{\sqrt r}\left(b_4^2\dd s^2_{\mathrm{AdS}_4}+  b_2^2 b_5^2(\hat{y}^i Dx^i)^2 \right) + \sqrt{r} \left(\frac{\dd r^2}{b_4^2} + b_2(\delta_{ij}- \hat{y}^i \hat{y}^j )D x^i D x^j+
	 \frac{b_1^2 b_5^2}{b_2 b_4^6} \dd s^2_{S^2}\right)\,.
\end{align}
(The explicit expression of $\tilde{f}_{21,0}(b_i)$ is not particularly interesting.) Now there are five directions multiplied by $h^{1/2}= \sqrt{r}$, so $p=4$. Again the linear behavior of $h$ can arise near the hole boundary of an O4.\footnote
{Another difference with (\ref{eq:o6}) is the behavior of $\psi$. In (\ref{eq:o6}), $\psi\to 0$; the pure $\Phi_\pm$ become $\mathrm{SU}(3)$-structure type, as for the ordinary O6-plane with $F_0=0$ (see for example \cite[Sec.~3]{saracco-t}). In (\ref{eq:o4}), $\psi\to \pi/2$; indeed the flat-space O4 solution is of $\mathrm{SU}(2)$-structure type, as can be seen by T-dualizing the O6 twice
. Notice that in IIA an AdS$_4$ solution cannot have $\mathrm{SU}(2)$ structure everywhere \cite{bovy-lust-tsimpis,caviezel-kors-lust-tsimpis-zagermann}. 
} 
In principle it could arise by expanding either $\dd s^2_\mathrm{O4}= h^{-1/2}\dd s^2_\parallel + h^{1/2} (\dd r^2 + r^2 \dd s^2_{\mathbb{R}^2 \times S^2})$ or $\dd s^2_\mathrm{smO4}= h^{-1/2}\dd s^2_\parallel + h^{1/2} (\dd r^2 + r^2 \dd s^2_{S^2} + \dd s^2_{\mathbb{R}^2})$. The first  possibility represents an O4 sitting at the tip of the cone $C(\mathbb{R}^2\times S^2)$, which is a badly singular space, unclear to make sense even in string theory. The second represents an O4 smeared along the $\dd s^2_{\mathbb{R}^2}= (\delta_{ij}- \hat{y}^i \hat{y}^j )D x^i D x^j$ directions, and appears to be the most likely interpretation. (For more details see Sec.~4.1.1 of \cite{passias-prins-t}, where such singularities appeared often.) As mentioned earlier, smearing an O-plane is of dubious physical validity. Moreover, this particular smearing is even more formal than usual: it seems to occur in the direction $\hat{y}^i \dd x^i$, which depends on the transverse coordinates. (Another way of seeing this problem is that the form along which one would measure O4 charge, $\omega_{2,1}\wedge \omega_{2,2}$, is not closed.)

\subparagraph{Other singularities.}

Several other attractors exist. The behavior we call $\sigma$ in Table \ref{tab:bc} is quite common; it does not obviously match an O-plane or D-brane singularity. Another, less common one has $\rho_1\sim 	\rho_2\sim r^{1/2}$, $\rho_3\sim r^{-1}$, $\tan \psi\sim r^{-1/2}$, with the other variables going to constants.


\subsection{Matching with the perturbative solution}

\label{sub:pertgs}

As we have seen in Section \ref{sub:bc}, there is a rich variety of local behaviors allowed by the system \eqref{eq:systemFinal} near a boundary.
A natural question to ask is then which of those can arise in the completion of the perturbative solution in \cite{marchesano-palti-quirant-t},
even though from the previous analysis we have indications that a completely smooth boundary seems to be forbidden.

The solution in \cite{marchesano-palti-quirant-t} was obtained perturbatively in $g_s \ll 1$ in the limit (\ref{eq:lim}), with the additional Ansatz
\begin{equation}
	\psi = g_s \psi_1 + O(g_s^3) \, ,\qquad \ee^\phi = g_s \ee^{3A_0 + g_s^2 \phi_2+ O(g_s^4)} \, ,\qquad
	 \ee^A = \ee^{A_0 + g_s^2 A_2+ O(g_s^4)}\,.
\end{equation}
On a general Calabi--Yau, the solution can be summarized as 
\begin{align}\label{eq:dgkt-approx}
	&J_\psi\sim J_\mathrm{CY} \, ,\qquad \omega = -\frac\ii{2 \psi_1} \bar v_1 \cdot \Omega\, ,\qquad 
	\ee^{A_0}\sim 1-g_s \varphi\,,\\
	\nonumber&\mathrm{Im} \Omega \sim (1+ g_s \varphi ) \mathrm{Im} \Omega_\mathrm{CY} - g_s *_\mathrm{CY} K \, ,\qquad 
	\mathrm{Re} \Omega \sim (1- g_s \varphi) \mathrm{Re} \Omega_\mathrm{CY} + g_s  K \,,\\
	\nonumber & 
	v\sim \frac12 g_s\ee^{A_0} \partial_\mathrm{CY} \tilde f + O(g_s^3)\, ,\qquad \Delta_\mathrm{CY}\tilde f= 8 g_s F_0 \varphi \, ,\qquad
	\Delta_\mathrm{CY} K = 2g_s F_0 \mathrm{Re} \Omega_\mathrm{CY}\,
\end{align}
One also obtains $3 A_2 - \phi_2 = \frac12 \psi_1^2-\frac15 F_0 f_\star$ for the subleading order.

This general solution was also made more explicit by taking the Calabi--Yau to be $T^6/\mathbb{Z}_2\times \mathbb{Z}_2$. As in the previous subsections, the coordinates are called $x^i$, $y^i$, and the O6 involution maps $(x^i,y^i)\to (x^i,-y^i)$; but all coordinates are now periodically identified, with unit period.
The generators of the $\mathbb{Z}_2 \times \mathbb{Z}_2$ can be taken to be $(x^1,x^2,x^3,y^1,y^2,y^3)\to (-x^1,-x^2,x^3,-y^1,-y^2,y^3)$ and $\to (-x^1,x^2,-x^3,-y^1,y^2,-y^3)$.
Because of this discrete identification, besides the O6-plane at $\{y^1=y^2=y^3=0\}$ we also have three more, at $\{x^1=x^2=y^3=0\}$, $\{x^1=y^2=x^3=0\}$ and $\{y^1=x^2=x^3=0\}$. 

The data in (\ref{eq:dgkt-approx}) are given in \cite[Sec.~6.2]{marchesano-palti-quirant-t} for this case, but we are interested in the behavior near one of these O6-planes, which without loss of generality can be taken to be the one at $\{y^1=y^2=y^3=0\}$; in other words we take a limit where $y^i\ll x^j$, $\forall i,\,j$. We specify the constants in \cite[Sec.~6.2]{marchesano-palti-quirant-t} as $qF_0=4$, $h=L^{3}$; moreover we need to rescale $(x^i,y^i)\to L (x^i,y^i)$. This leads to 
\begin{align}\label{eq:dgkt-approx2}
		\nonumber & J_\mathrm{CY} = \dd x^i \wedge \dd y^i \, ,\qquad \Omega_\mathrm{CY}= (\dd x^1+\ii \dd y^1) \wedge (\dd x^2+\ii \dd y^2) \wedge (\dd x^3+\ii \dd y^3) \, ,\\
		\nonumber & v= \frac{g_s F_0}{\pi r^3} Z^{-1/2} y^i (-\ii Z^{-1/4}\dd x^i + Z^{1/4} \dd y^i) \, ,\qquad j= J_0 -\frac{\ii}{2 \psi_1^2} v \wedge \bar v = \omega_{2,4}+\frac1{r^2} \omega_{2,3}\,\\
		& \omega= -\frac{\ii}r y^1(Z^{-1/4}\dd x^2 +\ii Z^{1/4} \dd y^2) \wedge (Z^{-1/4}\dd x^2 +\ii Z^{1/4} \dd y^2) + \mathrm{cycl}.
		\\ 
		\nonumber &\quad= -\frac{\ii}r \left(\frac12 Z^{-1/2} \omega_{2,2} + \ii \omega_{2,0} - \frac12 Z^{1/2} \omega_{2,1}\right) \, ;\\
		\nonumber & \ee^{A_0}= Z^{-1/4} \, ,\qquad \psi_1 = \frac{g_s F_0}{\pi r^2} Z^{-1/2}
		\, ,\qquad f_\star\sim -\frac{2g_s F_0}{\pi r}\,. 
\end{align}
$Z= 1-g_s/r$, $r=\sqrt{y^i y^i}$ is the usual harmonic function for the O6 in flat space.

Comparing the quantities in \eqref{eq:dgkt-approx2} with our definitions in Section \ref{sub:sym}, we extract the leading behaviors of the functions appearing in the system \eqref{eq:systemFinal}, obtaining also that $c_1$ has to be of order $g_s$ or smaller.

\begin{figure}
    \centering
    \begin{subfigure}[t]{0.48\textwidth}
        \centering
        \includegraphics[width=\linewidth]{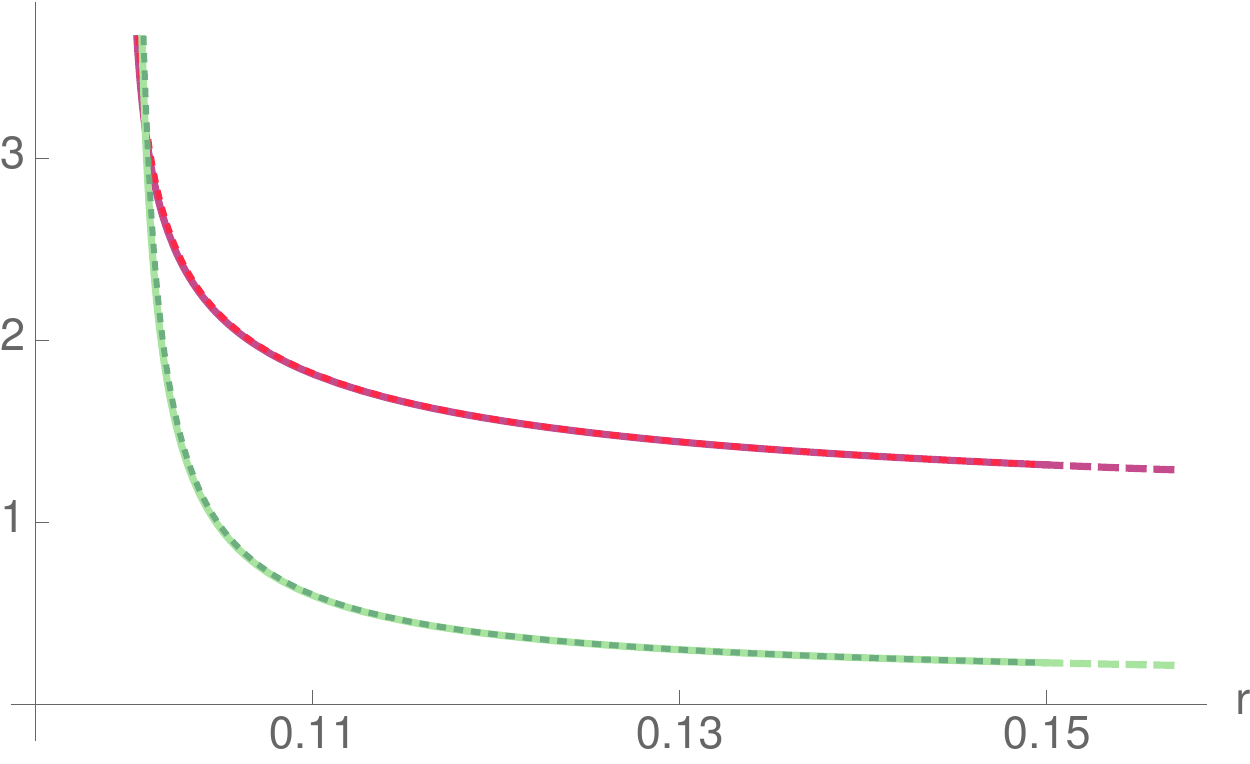} 
        \caption{$n_0=10^{-10}$ }\label{fig:numerics-n0tiny}
    \end{subfigure}\hfill
    \begin{subfigure}[t]{0.48\textwidth}
        \centering
        \includegraphics[width=\linewidth]{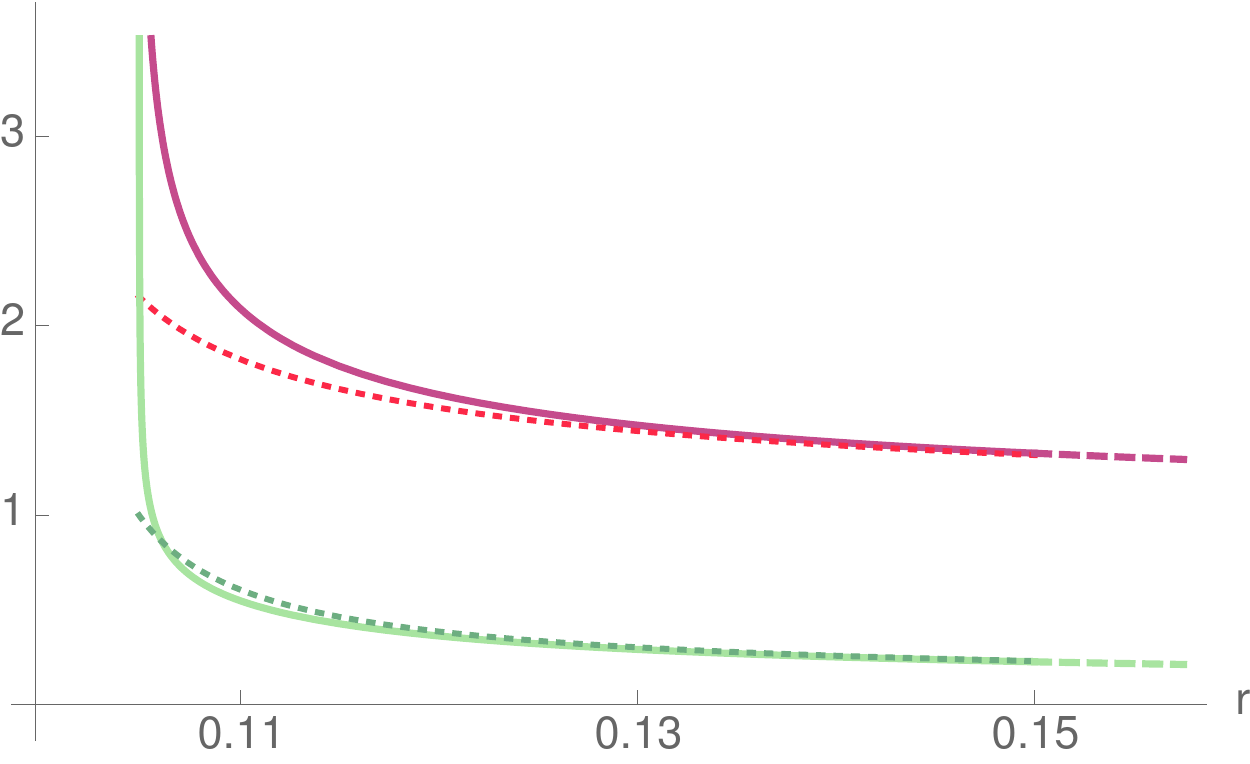} 
        \caption{$n_0=1$} \label{fig:numerics-n01}
    \end{subfigure}
    
	\caption{\small Numerical solutions obtained by imposing boundary conditions at $r = 2$, with $g_s = .1$, $c_1 =0$, for different values of $n_0$.
	In solid red (green) we show the behavior of $e^A$ ($e^\phi$) compared to the behavior of the perturbative solution (dotted).
	For $n_0 = 10^{-10}$ the functions are indistinguishable, but for $n_0 = 1$ the functions deviate from the perturbative ones, approaching the O4 singularity \eqref{eq:o4}.}
	  \label{fig:numerics1}
	\end{figure}

We can now evaluate the perturbative solution at an $r\gg g_s$, where it should be reliable,
and we use it to extract the initial data to compute a numerical solution of the system \eqref{eq:systemFinal}, evolving towards the locus where the O6 would be expected to be.
As a check of this numerical method, we first performed the analysis for a tiny $n_0 \equiv 2\pi F_0\ll 1$, a limit in which the solution should reproduce the massless one with high accuracy.
This works out correctly, as it can be seen from the example in Figure \ref{fig:numerics-n0tiny}, where the numerical solution is indistinguishable from the perturbative one describing a massless O6.
This check is non-trivial from the point of view of the system \eqref{eq:systemFinal}, since the O6 boundary condition appeared to be fine-tuned and not a generic attractor, as we stressed in Sec.~\ref{sub:bc}.

We then increased $n_0$ up to 1, keeping $g_s$ and all the other perturbative parameters fixed, and we obtained that near the end the solution starts deviating from the perturbative one (\ref{eq:dgkt-approx2}).
Surprisingly, however, it does not remain of the O6 kind, but it switches to the partially smeared O4 type in Table \ref{tab:bc}, with the functions locally behaving as in \eqref{eq:o4}.
An example is shown in Figure \ref{fig:numerics-n01}.

Increasing $n_0$ even more, the singular behavior changes again approaching the class $\sigma$ in Table \ref{tab:bc}.
To make sure this behavior is generic in the space of the perturbative parameters, we ran $\sim 10^5$ numerical evolutions with randomly selected parameters.
In all the cases we only obtained either smeared-O4 singularities or $\sigma$-type singularities.
For all the latter, we have checked that decreasing $n_0$, without making it tiny, always resulted in smeared O4 solutions, showing that the behavior discussed above as function of $n_0$ is generic. Finally, for  $g_s$ extremely small ($<10^{-3}$) the system evolves towards another class of singularities at $r\ll1$, briefly mentioned at the end of Section \ref{sub:bc}, whose meaning is not clear.

The solutions of \cite{junghans-dgkt,marchesano-palti-quirant-t} are approximate, and are expected to receive small corrections at the next level in $g_s$. To account for this, we again performed $\sim 10^5$ evolutions, adding up to 10\% random noise to the boundary conditions determined from (\ref{eq:dgkt-approx2}). Again the O4 singularity was by far the most common, but sporadically other non-smooth behaviors also appeared (in about $< 0.05\%$ of the cases). Among these, we found a small number of candidate O6 solutions. This agrees with our observation in Sec.~\ref{sub:bc} that these require fine tuning. So it is possible that some perturbations of (\ref{eq:dgkt-approx2}) can be glued to an O6 behavior, as one might have expected. We stress once again that the present analysis is only local; in particular we have not imposed the $F_4$ flux quantization. 

In an even smaller number of cases, the evolution stops at a point where $\rho_3=\psi=0$ and all the other variables are finite and non-zero; at this point the metric remains finite, but as discussed in Sec.~\ref{sub:bc} this solution is not smooth when glued to a second copy.


\subsection{Summary} 
\label{sub:summary}

This section was motivated by the hope that the O6-planes that originally motivated the AdS$_4\times \mathrm{CY}_6$ solution of \cite{dewolfe-giryavets-kachru-taylor} could be replaced by smooth loci. This phenomenon arose generically in \cite{saracco-t} in a similar context, for a local solution obtained by deforming the O6-plane by Romans mass $F_0$,  which however could not be made compact. Moreover, the general bounds of Sec.~\ref{sec:be} left this possibility open.

We found in Sec.~\ref{sub:bc} that there is no such a local behavior. Among the attractor boundary conditions for the supersymmetry equations, there is one solution that closely resembles that in \cite[Sec.~7]{saracco-t}, but if we glue it to a second copy of itself to avoid a boundary, we find in fact an angular point in one of the metric coefficients.

We also found in Sec.~\ref{sub:pertgs} that it is quite hard to glue the solution in \cite{marchesano-palti-quirant-t} even to the local O6 singularity for the system. Generically the solution is instead attracted to the formal partially smeared O4 singularity (\ref{eq:o4}). This state of affairs might perhaps be ameliorated with more precise numerical work, given that the O6 boundary condition is not an attractor for the general supersymmetry system. 

But as already pointed out in \cite{junghans-dgkt}, achieving this gluing would be in any case only a marginal improvement over the approximate solutions in \cite{marchesano-palti-quirant-t}. For $g_s \ll 1$, those could be trusted in most of the internal space; their approximation breaks down at a distance of order a few $g_s$ around the O6-planes. The full system studied in this section can be trusted in a region that is smaller, but also of radius $g_s$ in string units. The real motivation for the study in this section was to avoid a singularity altogether; that would have boosted our confidence in the solution enormously, but as we saw it cannot be achieved.



\section*{Acknowledgements}

We would like to thank N.~De Ponti, S.~Meda, A.~Mondino, E.~Silverstein, G.~Torroba and especially A.~Hassannezhad for interesting discussions and correspondence. We also thank C.~C\'ordova for collaboration during the initial phase of this project. GBDL is supported in part by the Simons Foundation Origins of the Universe Initiative (modern inflationary cosmology collaboration) and by a Simons Investigator award. AT is supported in part by INFN and by MIUR-PRIN contract 2017CC72MK003. 

\bibliography{at}

\providecommand{\href}[2]{#2}\begin{thebibliography}{10}

\bibitem{kklt}
S.~Kachru, R.~Kallosh, A.~Linde, and S.~P. Trivedi, ``{De Sitter} vacua in
  string theory,'' {\em Phys. Rev.} {\bf D68} (2003) 046005,
\href{http://arXiv.org/abs/hep-th/0301240}{{\tt hep-th/0301240}}.

\bibitem{palti-vafa-weigand}
E.~Palti, C.~Vafa, and T.~Weigand, ``{Supersymmetric Protection and the
  Swampland},'' {\em JHEP} {\bf 06} (2020) 168,
\href{http://arXiv.org/abs/2003.10452}{{\tt 2003.10452}}.

\bibitem{dewolfe-giryavets-kachru-taylor}
O.~DeWolfe, A.~Giryavets, S.~Kachru, and W.~Taylor, ``Type {IIA} moduli
  stabilization,'' {\em JHEP} {\bf 07} (2005) 066,
\href{http://arXiv.org/abs/hep-th/0505160}{{\tt hep-th/0505160}}.

\bibitem{camara-font-ibanez}
P.~G. Camara, A.~Font, and L.~E. {Ib\'a\~nez}, ``Fluxes, moduli fixing and
  {MSSM-like} vacua in a simple {IIA} orientifold,'' {\em JHEP} {\bf 09} (2005)
  013,
\href{http://arXiv.org/abs/hep-th/0506066}{{\tt hep-th/0506066}}.

\bibitem{polchinski-silverstein}
J.~Polchinski and E.~Silverstein, ``{Dual Purpose Landscaping Tools: Small
  Extra Dimensions in AdS/CFT},'' in {\em Strings, gauge fields, and the
  geometry behind: The legacy of Maximilian Kreuzer}, A.~Rebhan, L.~Katzarkov,
  J.~Knapp, R.~Rashkov, and E.~Scheidegger, eds., pp.~365--390.
\newblock 2009.
\newblock
\href{http://arXiv.org/abs/0908.0756}{{\tt 0908.0756}}.
\newblock

\bibitem{ooguri-vafa-swamp}
H.~Ooguri and C.~Vafa, ``{On the Geometry of the String Landscape and the
  Swampland},'' {\em Nucl. Phys.} {\bf B766} (2007) 21--33,
\href{http://arXiv.org/abs/hep-th/0605264}{{\tt hep-th/0605264}}.

\bibitem{lust-palti-vafa}
D.~{L\"ust}, E.~Palti, and C.~Vafa, ``{AdS and the Swampland},'' {\em Phys.
  Lett.} {\bf B797} (2019) 134867,
\href{http://arXiv.org/abs/1906.05225}{{\tt 1906.05225}}.

\bibitem{gautason-schillo-vanriet-williams}
F.~F. Gautason, M.~Schillo, T.~Van~Riet, and M.~Williams, ``{Remarks on scale
  separation in flux vacua},'' {\em JHEP} {\bf 03} (2016) 061,
\href{http://arXiv.org/abs/1512.00457}{{\tt 1512.00457}}.

\bibitem{apruzzi-deluca-gnecchi-lomonaco-t}
F.~Apruzzi, G.~Bruno De~Luca, A.~Gnecchi, G.~Lo~Monaco, and A.~Tomasiello,
  ``{On AdS$_{7}$ stability},'' {\em JHEP} {\bf 07} (2020) 033,
\href{http://arXiv.org/abs/1912.13491}{{\tt 1912.13491}}.

\bibitem{bachas-estes}
C.~Bachas and J.~Estes, ``{Spin-2 spectrum of defect theories},'' {\em JHEP}
  {\bf 06} (2011) 005,
\href{http://arXiv.org/abs/1103.2800}{{\tt 1103.2800}}.

\bibitem{csaki-erlich-hollowood-shirman}
C.~Csaki, J.~Erlich, T.~J. Hollowood, and Y.~Shirman, ``{Universal aspects of
  gravity localized on thick branes},'' {\em Nucl. Phys.} {\bf B581} (2000)
  309--338,
\href{http://arXiv.org/abs/hep-th/0001033}{{\tt hep-th/0001033}}.

\bibitem{hassannezhad}
A.~Hassannezhad, ``Eigenvalues of perturbed laplace operators on compact
  manifolds,'' {\em Pacific Journal of Mathematics} {\bf 264} (2013), no.~2,
  333--354.

\bibitem{setti-eigenvalues}
A.~G. Setti, ``Eigenvalue estimates for the weighted laplacian on a riemannian
  manifold,'' {\em Rendiconti del Seminario Matematico della Universit\`a di
  Padova} {\bf 100} (1998) 27--55.

\bibitem{saracco-t}
F.~Saracco and A.~Tomasiello, ``{Localized O6-plane solutions with Romans
  mass},'' {\em JHEP} {\bf 1207} (2012) 077,
\href{http://arXiv.org/abs/1201.5378}{{\tt 1201.5378}}.

\bibitem{junghans-dgkt}
D.~Junghans, ``{O-plane Backreaction and Scale Separation in Type IIA Flux
  Vacua},'' {\em Fortsch. Phys.} {\bf 68} (2020), no.~6, 2000040,
\href{http://arXiv.org/abs/2003.06274}{{\tt 2003.06274}}.

\bibitem{marchesano-palti-quirant-t}
F.~Marchesano, E.~Palti, J.~Quirant, and A.~Tomasiello, ``{On supersymmetric
  AdS$_{4}$ orientifold vacua},'' {\em JHEP} {\bf 08} (2020) 087,
\href{http://arXiv.org/abs/2003.13578}{{\tt 2003.13578}}.

\bibitem{afrt}
F.~Apruzzi, M.~Fazzi, D.~Rosa, and A.~Tomasiello, ``{All AdS$_7$ solutions of
  type II supergravity},'' {\em JHEP} {\bf 1404} (2014) 064,
\href{http://arXiv.org/abs/1309.2949}{{\tt 1309.2949}}.

\bibitem{cremonesi-t}
S.~Cremonesi and A.~Tomasiello, ``{6d holographic anomaly match as a continuum
  limit},'' {\em JHEP} {\bf 05} (2016) 031,
\href{http://arXiv.org/abs/1512.02225}{{\tt 1512.02225}}.

\bibitem{apruzzi-fazzi}
F.~Apruzzi and M.~Fazzi, ``{AdS$_{7}$/CFT$_{6}$ with orientifolds},'' {\em
  JHEP} {\bf 01} (2018) 124,
\href{http://arXiv.org/abs/1712.03235}{{\tt 1712.03235}}.

\bibitem{passias-prins-t}
A.~Passias, D.~Prins, and A.~Tomasiello, ``{A massive class of $\mathcal{N} =
  2$ AdS$_4$ IIA solutions},'' {\em JHEP} {\bf 10} (2018) 071,
\href{http://arXiv.org/abs/1805.03661}{{\tt 1805.03661}}.

\bibitem{acharya-benini-valandro}
B.~S. Acharya, F.~Benini, and R.~Valandro, ``Fixing moduli in exact type {IIA}
  flux vacua,'' {\em JHEP} {\bf 02} (2007) 018,
\href{http://arXiv.org/abs/hep-th/0607223}{{\tt hep-th/0607223}}.

\bibitem{lichnerowicz-bound}
A.~Lichnerowicz, ``G{\'e}om{\'e}trie des groupes de transformations,''.

\bibitem{li-yau}
P.~Li and S.~T. Yau, ``Estimates of eigenvalues of a compact {R}iemannian
  manifold,'' in {\em Geometry of the {L}aplace operator ({P}roc. {S}ympos.
  {P}ure {M}ath., {U}niv. {H}awaii, {H}onolulu, {H}awaii, 1979)}, Proc. Sympos.
  Pure Math., XXXVI, pp.~205--239.
\newblock Amer. Math. Soc., Providence, R.I., 1980.

\bibitem{douglas-kallosh}
M.~R. Douglas and R.~Kallosh, ``{Compactification on negatively curved
  manifolds},''
\href{http://arXiv.org/abs/1001.4008}{{\tt 1001.4008}}.

\bibitem{maldacena-nunez}
J.~M. Maldacena and C.~N{\'u}{\~n}ez, ``Supergravity description of field
  theories on curved manifolds and a no-go theorem,'' {\em Int. J. Mod. Phys.}
  {\bf A16} (2001) 822--855,
\href{http://arXiv.org/abs/hep-th/0007018}{{\tt hep-th/0007018}}.

\bibitem{cordova-deluca-t-dso6}
C.~C\'{o}rdova, G.~B. De~Luca, and A.~Tomasiello, ``{New de Sitter Solutions in
  Ten Dimensions and Orientifold Singularities},'' {\em JHEP} {\bf 08} (2020)
  093, \href{http://arXiv.org/abs/1911.04498}{{\tt 1911.04498}}.

\bibitem{cheng-bound}
S.-Y. Cheng, ``Eigenvalue comparison theorems and its geometric applications,''
  {\em Mathematische Zeitschrift} {\bf 143} (1975), no.~3, 289--297.

\bibitem{klebanov-pufu-rocha}
I.~R. Klebanov, S.~S. Pufu, and F.~D. Rocha, ``{The Squashed, Stretched, and
  Warped Gets Perturbed},'' {\em JHEP} {\bf 06} (2009) 019,
\href{http://arXiv.org/abs/0904.1009}{{\tt 0904.1009}}.

\bibitem{ahn-woo}
C.~Ahn and K.~Woo, ``{Perturbing Around A Warped Product Of AdS$_4$ and
  Seven-Ellipsoid},'' {\em JHEP} {\bf 08} (2009) 065,
\href{http://arXiv.org/abs/0907.0969}{{\tt 0907.0969}}.

\bibitem{richard-terrisse-tsimpis}
J.-M. Richard, R.~Terrisse, and D.~Tsimpis, ``{On the spin-2 Kaluza--Klein
  spectrum of $ {\mathrm{AdS}}_4\times
  {S}^2\left({\mathrm{\mathcal{B}}}_4\right) $},'' {\em JHEP} {\bf 12} (2014)
  144,
\href{http://arXiv.org/abs/1410.4669}{{\tt 1410.4669}}.

\bibitem{passias-t}
A.~Passias and A.~Tomasiello, ``{Spin-2 spectrum of six-dimensional field
  theories},'' {\em JHEP} {\bf 12} (2016) 050,
\href{http://arXiv.org/abs/1604.04286}{{\tt 1604.04286}}.

\bibitem{pang-rong-varela}
Y.~Pang, J.~Rong, and O.~Varela, ``{Spectrum universality properties of
  holographic Chern--Simons theories},'' {\em JHEP} {\bf 01} (2018) 061,
\href{http://arXiv.org/abs/1711.07781}{{\tt 1711.07781}}.

\bibitem{gutperle-uhlemann-spin2}
M.~Gutperle, C.~F. Uhlemann, and O.~Varela, ``{Massive spin 2 excitations in
  $AdS_6\times S^2$ warped spacetimes},'' {\em JHEP} {\bf 07} (2018) 091,
  \href{http://arXiv.org/abs/1805.11914}{{\tt 1805.11914}}.

\bibitem{passias-richmond}
A.~Passias and P.~Richmond, ``{Perturbing AdS$_6 \times_w S^4$: linearised
  equations and spin-2 spectrum},'' {\em JHEP} {\bf 07} (2018) 058,
\href{http://arXiv.org/abs/1804.09728}{{\tt 1804.09728}}.

\bibitem{malek-samtleben-kk}
E.~Malek and H.~Samtleben, ``{Kaluza--Klein Spectrometry for Supergravity},''
  {\em Phys. Rev. Lett.} {\bf 124} (2020), no.~10, 101601,
\href{http://arXiv.org/abs/1911.12640}{{\tt 1911.12640}}.

\bibitem{malek-nicolai-samtleben}
E.~Malek, H.~Nicolai, and H.~Samtleben, ``{Tachyonic Kaluza--Klein modes and
  the AdS swampland conjecture},'' {\em JHEP} {\bf 08} (2020) 159,
\href{http://arXiv.org/abs/2005.07713}{{\tt 2005.07713}}.

\bibitem{andriot-lucenagomez}
D.~Andriot and G.~Lucena~{G\'omez}, ``{Signatures of extra dimensions in
  gravitational waves},'' {\em JCAP} {\bf 1706} (2017) 048,
  \href{http://arXiv.org/abs/1704.07392}{{\tt 1704.07392}}.
[Erratum: JCAP1905,E01(2019)].

\bibitem{andriot-tsimpis}
D.~Andriot and D.~Tsimpis, ``{Gravitational waves in warped
  compactifications},'' {\em JHEP} {\bf 06} (2020) 100,
\href{http://arXiv.org/abs/1911.01444}{{\tt 1911.01444}}.

\bibitem{andriot-marconnet-tsimpis}
D.~Andriot, P.~Marconnet, and D.~Tsimpis, ``{Warp factor and the gravitational
  wave spectrum},''
\href{http://arXiv.org/abs/2103.09240}{{\tt 2103.09240}}.

\bibitem{giddings-maharana}
S.~B. Giddings and A.~Maharana, ``{Dynamics of warped compactifications and the
  shape of the warped landscape},'' {\em Phys. Rev.} {\bf D73} (2006) 126003,
\href{http://arXiv.org/abs/hep-th/0507158}{{\tt hep-th/0507158}}.

\bibitem{douglas-shiu-torroba-underwood}
G.~Shiu, G.~Torroba, B.~Underwood, and M.~R. Douglas, ``{Dynamics of Warped
  Flux Compactifications},'' {\em JHEP} {\bf 06} (2008) 024,
\href{http://arXiv.org/abs/0803.3068}{{\tt 0803.3068}}.

\bibitem{tsimpis-scale}
D.~Tsimpis, ``{Supersymmetric AdS vacua and separation of scales},'' {\em JHEP}
  {\bf 08} (2012) 142,
\href{http://arXiv.org/abs/1206.5900}{{\tt 1206.5900}}.

\bibitem{bakry-emery}
D.~Bakry and M.~\'Emery, ``Diffusions hypercontractives,'' {\em S\'eminaire de
  probabilit\'es de Strasbourg} {\bf 19} (1985) 177--206.

\bibitem{witten-morse}
E.~Witten, ``{Supersymmetry and Morse theory},'' {\em J. Diff. Geom.} {\bf 17}
  (1982), no.~4, 661--692.

\bibitem{gibbons-nogo}
G.~W. Gibbons, ``{Aspects of Supergravity Theories},''. Lectures at San Feliu
  de Guixols, Spain, June, 1984.

\bibitem{dewit-smit-haridass}
B.~de~Wit, D.~J. Smit, and N.~D. Hari~Dass, ``{Residual Supersymmetry of
  Compactified $D=10$ Supergravity},'' {\em Nucl. Phys.} {\bf B283} (1987)
165.

\bibitem{petrini-solard-vanriet}
M.~Petrini, G.~Solard, and T.~Van~Riet, ``{AdS vacua with scale separation from
  IIB supergravity},'' {\em JHEP} {\bf 1311} (2013) 010,
\href{http://arXiv.org/abs/1308.1265}{{\tt 1308.1265}}.

\bibitem{arkanihamed-dubovsky-nicolis-villadoro}
N.~Arkani-Hamed, S.~Dubovsky, A.~Nicolis, and G.~Villadoro, ``{Quantum Horizons
  of the Standard Model Landscape},'' {\em JHEP} {\bf 06} (2007) 078,
\href{http://arXiv.org/abs/hep-th/0703067}{{\tt hep-th/0703067}}.

\bibitem{toappear-casimir}
G.~B. De~Luca, E.~Silverstein, A.~Tomasiello, and G.~Torroba. Work in progress.

\bibitem{deluca-deponti-mondino-t}
G.~B. De~Luca, N.~De~Ponti, A.~Mondino, and A.~Tomasiello, ``A cheeger bound on
  light spin-two fields,''. To appear.

\bibitem{deluca-deponti-mondino-t2}
G.~B. De~Luca, N.~De~Ponti, A.~Mondino, and A.~Tomasiello. Work in progress.

\bibitem{buratti-calderon-mininno-uranga}
G.~Buratti, J.~Calderon, A.~Mininno, and A.~M. Uranga, ``{Discrete Symmetries,
  Weak Coupling Conjecture and Scale Separation in AdS Vacua},'' {\em JHEP}
  {\bf 06} (2020) 083,
\href{http://arXiv.org/abs/2003.09740}{{\tt 2003.09740}}.

\bibitem{wei-wylie}
G.~Wei and W.~Wylie, ``Comparison geometry for the bakry-emery ricci tensor,''
  {\em Journal of differential geometry} {\bf 83} (2009), no.~2, 337--405.

\bibitem{charalambous-lu-rowlett}
N.~Charalambous, Z.~Lu, and J.~Rowlett, ``Eigenvalue estimates on
  bakry-–{\'emery} manifolds,'' {\em Springer Proceedings in Mathematics \&
  Statistics} (2015) 45–61.

\bibitem{wu2013upper}
J.-Y. Wu, ``Upper bounds on the first eigenvalue for a diffusion operator via
  bakry--{\'e}mery ricci curvature ii,'' {\em Results in Mathematics} {\bf 63}
  (2013), no.~3, 1079--1094.

\bibitem{saracco-t-torroba}
F.~Saracco, A.~Tomasiello, and G.~Torroba, ``{Topological resolution of gauge
  theory singularities},'' {\em Phys.Rev.} {\bf D88} (2013) 045018,
\href{http://arXiv.org/abs/1305.2929}{{\tt 1305.2929}}.

\bibitem{mcorist-sethi}
J.~McOrist and S.~Sethi, ``{M-theory and Type IIA Flux Compactifications},''
  {\em JHEP} {\bf 12} (2012) 122,
\href{http://arXiv.org/abs/1208.0261}{{\tt 1208.0261}}.

\bibitem{rota-t}
A.~Rota and A.~Tomasiello, ``{AdS$\_{4}$ compactifications of AdS$\_{7}$
  solutions in type II supergravity},'' {\em JHEP} {\bf 07} (2015) 076,
\href{http://arXiv.org/abs/1502.06622}{{\tt 1502.06622}}.

\bibitem{berkovits}
N.~Berkovits, ``{Super-Poincar{\'e} covariant quantization of the
  superstring},'' {\em JHEP} {\bf 04} (2000) 018,
\href{http://arXiv.org/abs/hep-th/0001035}{{\tt hep-th/0001035}}.

\bibitem{benichou-policastro-troost}
R.~Benichou, G.~Policastro, and J.~Troost, ``{T-duality in Ramond-Ramond
  backgrounds},'' {\em Phys. Lett.} {\bf B661} (2008) 192--195,
\href{http://arXiv.org/abs/0801.1785}{{\tt 0801.1785}}.

\bibitem{gmpt2}
M.~Gra{\~n}a, R.~Minasian, M.~Petrini, and A.~Tomasiello, ``Generalized
  structures of {${\cal N}=1$} vacua,'' {\em JHEP} {\bf 11} (2005) 020,
\href{http://arXiv.org/abs/hep-th/0505212}{{\tt hep-th/0505212}}.

\bibitem{gmpt3}
M.~Gra{\~n}a, R.~Minasian, M.~Petrini, and A.~Tomasiello, ``A scan for new
  {${\cal N}=1$} vacua on twisted tori,'' {\em JHEP} {\bf 05} (2007) 031,
\href{http://arXiv.org/abs/hep-th/0609124}{{\tt hep-th/0609124}}.

\bibitem{t-reform}
A.~Tomasiello, ``{Reformulating Supersymmetry with a Generalized {Dolbeault}
  Operator},'' {\em JHEP} {\bf 02} (2008) 010,
\href{http://arXiv.org/abs/arXiv:0704.2613 [hep-th]}{{\tt arXiv:0704.2613
  [hep-th]}}.

\bibitem{lust-tsimpis}
D.~{L\"ust} and D.~Tsimpis, ``Supersymmetric {AdS$_4$} compactifications of
  {IIA} supergravity,'' {\em JHEP} {\bf 02} (2005) 027,
\href{http://arXiv.org/abs/hep-th/0412250}{{\tt hep-th/0412250}}.

\bibitem{bovy-lust-tsimpis}
J.~Bovy, D.~Lust, and D.~Tsimpis, ``{${\mathcal N} = 1,2$ supersymmetric vacua
  of IIA supergravity and SU(2) structures},'' {\em JHEP} {\bf 08} (2005) 056,
\href{http://arXiv.org/abs/hep-th/0506160}{{\tt hep-th/0506160}}.

\bibitem{caviezel-kors-lust-tsimpis-zagermann}
C.~Caviezel, P.~Koerber, S.~{K\"ors}, D.~{L\"ust}, D.~Tsimpis, and
  M.~Zagermann, ``{The effective theory of type IIA AdS$_4$ compactifications
  on nilmanifolds and cosets},''
\href{http://arXiv.org/abs/0806.3458}{{\tt 0806.3458}}.

\end{thebibliography}
\bibliographystyle{at}

\end{document}